\documentclass[twocolumn,superscriptaddress,
amsmath,amssymb,aps,pr,nofootinbib]{revtex4-1}

\usepackage{graphicx}
\usepackage{physics}
\usepackage{comment}
\usepackage{color}
\usepackage{dcolumn}
\usepackage[normalem]{ulem}
\usepackage{bm}
\usepackage{multirow}
\usepackage[colorlinks,urlcolor=blue,citecolor=blue,linkcolor=blue]{hyperref}

\newcommand{\vect}[1]{{\mathbf #1}}
\newcommand{\vegr}[1]{{\boldsymbol #1}}   
\newcommand{\kv}{{\bf k}}

\newcommand{\0}{{\bf 0}}
\renewcommand{\r}{{\bf r}}
\newcommand{\rhov}{{\bm \rho}}

\newcommand{\sch}{Schr{\"o}dinger }

\begin{document}

\title{Magnetic-field driven hybridization of heavy- and light-hole Rydberg excitons in GaAs quantum wells}

\author{David de la Fuente Pico}
\affiliation{Departamento de F\'isica Te\'orica de la Materia
  Condensada, Universidad
  Aut\'onoma de Madrid, Madrid 28049, Spain}
\affiliation{Condensed Matter Physics Center (IFIMAC), Universidad Autónoma de Madrid, 28049 Madrid, Spain}

\author{Johannes B\"urger}
\affiliation{CNR Nanotec, Institute of Nanotechnology, via Monteroni, 73100 Lecce, Italy}

\author{Antonio Gianfrate}
\affiliation{CNR Nanotec, Institute of Nanotechnology, via Monteroni, 73100 Lecce, Italy}

\author{Jesper Levinsen}
\affiliation{School of Physics and Astronomy, Monash University, Victoria 3800, Australia}

\author{Meera M. Parish}
\affiliation{School of Physics and Astronomy, Monash University, Victoria 3800, Australia}

\author{Daniele Sanvitto}
\affiliation{CNR Nanotec, Institute of Nanotechnology, via Monteroni, 73100 Lecce, Italy}

\affiliation{Condensed Matter Physics Center (IFIMAC), Universidad Autónoma de Madrid, 28049 Madrid, Spain}

\author{Dario Ballarini}
\affiliation{CNR Nanotec, Institute of Nanotechnology, via Monteroni, 73100 Lecce, Italy}

\author{Francesca Maria Marchetti}
\affiliation{Departamento de F\'isica Te\'orica de la Materia
  Condensada, Universidad
  Aut\'onoma de Madrid, Madrid 28049, Spain}
\affiliation{Condensed Matter Physics Center (IFIMAC), Universidad Autónoma de Madrid, 28049 Madrid, Spain}  

\date{June 3, 2026}

\begin{abstract}
We present a combined theoretical and experimental study of ground and excited Rydberg exciton states in wide GaAs quantum wells exposed to a magnetic field in the Faraday geometry. We employ a multiband exciton model based on the Luttinger Hamiltonian, which captures valence-band mixing between heavy- and light-hole states induced by both the quantum well confinement and the magnetic field, and we develop an efficient numerical approach to solve for both ground- and excited-state excitons. The method treats Coulomb interactions, magnetic confinement, and band mixing on an equal footing, enabling a systematic characterization of exciton energies, oscillator strengths, and orbital composition. We show that band hybridization increases with magnetic field and is significantly more pronounced for higher excited states, where it sets in at lower fields and strongly modifies their properties. The theoretical predictions are validated by polarization-resolved magneto-reflectance measurements up to $9$~T on GaAs/Al$_{0.4}$Ga$_{0.6}$As quantum wells of $20$~nm width. We find excellent agreement for both the diamagnetic shift and Zeeman splitting of the ground state and the first four Rydberg excitons. Our results demonstrate that valence-band mixing plays a crucial role in determining the magnetic-field dependence of excited exciton states and must be properly included for a quantitative description of magneto-excitons in wide GaAs quantum wells.
\end{abstract}

\maketitle
\section{Introduction}
Magneto-optical spectroscopy has been widely employed to probe and control the properties of optical excitations in two-dimensional (2D) semiconductors~\cite{Miura_book2007}, including III–V~\cite{Tarucha_SSC1984,Bugajski_SSC1986,Rogers_PRB1986,Plaut_PRB1988,Shields_PRB1995} and II–VI~\cite{Warnock_PRB1985,Ivchenko_PRB1992,Kheng_PRL1993, Astakhov_PRB2002} quantum wells (QWs). A magnetic field applied in the Faraday geometry, i.e., perpendicular to the QW, enables the resolution and characterization of both ground and excited-state excitons. The resulting diamagnetic shift provides direct information on the exciton size and constituent carrier masses~\cite{Arora-Review_JAP2021}, while Zeeman splitting~\cite{Traynor_PRB1997, Chen_APS2006,Arora-Sandip_JAP2013,Grigoryev_PRB2016} provides a signature of its spin properties. At the same time, magnetic confinement enhances the exciton binding energy and oscillator strength, i.e., its coupling to light. This effect is particularly important for Rydberg states, whose oscillator strength rapidly decreases with increasing principal quantum number, making them otherwise difficult to access optically.

From a modeling perspective, magneto-optical measurements have played a central role in validating Wannier-based exciton models in, e.g., GaAs QWs, enabling the use of simplified effective-mass descriptions of excitons~\cite{Edelstein_PRB1989,Stafford_PRB1990,Yang_PRA1991}. 
It has been recognized early on that an accurate description of exciton properties requires the inclusion of valence band mixing induced by both quantum confinement and the magnetic field. 
This becomes particularly important in wider QWs, where exciton states associated with different valence subbands, the \emph{light-hole} and \emph{heavy-hole} excitons, approach each other in energy and can hybridize, an effect that is 
further enhanced by the magnetic field due to its in-plane confining effect. 
The properties of the magnetoexciton ground state have been extensively investigated within this framework~\cite{Yang_PRL1987,Fancilotto_SLattM1987,Bauer_AndoPRB1988,Bauer-Ando2_PRB1988,Potemski-Vinha_PRB1991, Traynor_PRB1997}.
More recently, accurate modeling including valence-band mixing effects has been employed to describe excited-subband  exciton states, which have become experimentally accessible owing to advances in sample quality~\cite{Bataev-Ignatiev-Efimov_PRB2022,Grigoryev_PRB2016,Grigoryev_JETP2023,Grigoryev-Chukeev_JAP2025}.
However, little attention has been devoted to the corresponding behavior of Rydberg exciton states. In this case, band-mixing effects have mainly been investigated at low magnetic fields~\cite{Vinha-Bauer_PRB1990}.

In this work, we theoretically investigate the ground and excited states of heavy- and light-hole excitons in wide GaAs QWs, and validate the results experimentally using an 18 nm sample.
We employ a multiband exciton model based on the Luttinger Hamiltonian, originally developed in the presence of a magnetic field by Bauer and Ando~\cite{Bauer_AndoPRB1988}. This approach provides the simplest effective-mass description that includes light- and heavy-hole valence-band mixing in magnetoexcitons and has proven essential for accurately describing the ground state exciton properties in intermediate and wide GaAs QWs~\cite{Bauer-Ando2_PRB1988,Traynor_PRB1997,Potemski-Vinha_PRB1991}.
We develop an efficient numerical method to solve these equations for both ground and excited Rydberg states. The in-plane degrees of freedom are treated by exact diagonalization in momentum space, combined with a Landé subtraction scheme to regularize the Coulomb singularity~\cite{Laird_PRB2022,deLaFuentePico_PRB2025,Kumar-Levinsen_PRB2025}, here adapted to finite-width QWs. This approach treats Coulomb interaction, valence-band mixing, and magnetic-field effects on equal footing. The out-of-plane degrees of freedom are described using an expansion over a finite basis of confined states.
We find that band hybridization increases with magnetic field and is more pronounced for higher excited states, where it sets in at lower fields.
We characterize both ground and excited exciton states according to their oscillator strength and their in-plane orbital symmetry.

Our theoretical modeling is validated by the excellent agreement with experimental results, without the use of any adjustable parameters apart from a slight reduction of the QW width. Exciton energies for magnetic fields up to $9$~T in the Faraday geometry are extracted from polarization-resolved reflectance measurements performed on a heterostructure sample containing 12 GaAs/Al$_{0.4}$Ga$_{0.6}$As QWs of $20$~nm width.
We find excellent agreement between theory and experiment for both the exciton diamagnetic shift and Zeeman splitting of the ground state as well as the first four excited Rydberg states. We show that properly accounting for band hybridization is essential in order to reproduce the magnetic-field dependence of both quantities.

This paper is organized as follows. In Sec.~\ref{sec:model}, we introduce the multiband Luttinger model for excitons in GaAs QWs exposed to a perpendicular magnetic field and derive the corresponding set of coupled \sch equations in momentum space.  Section~\ref{sec:results} presents our numerical scheme and results for the exciton energies, and an analysis of the exciton properties at zero (Sec.~\ref{eq:zero-field}) and finite magnetic field (Sec.~\ref{sec:results-MF}). We compare theory and experiment in Sec.~\ref{sec:experiments}, focusing on the magnetic-field dependence of the exciton diamagnetic shift and on the Zeeman splitting. Conclusions are drawn in Sec.~\ref{sec:conclusions}.

\section{Model}
\label{sec:model}
In this section, we introduce the microscopic model used to describe excitonic properties in finite-width GaAs quantum wells (QWs) exposed to a perpendicular magnetic field, including band-mixing effects induced by both quantum confinement and the applied field. 
We employ an effective-mass theory based on the Luttinger Hamiltonian formalism introduced in the seminal work of Bauer and Ando~\cite{Bauer_AndoPRB1988}, which we briefly rederive here for completeness. We then formulate the 
\sch equation in momentum space, a representation that proves advantageous for numerical solution, enabling the calculation of the full Rydberg series of bound excitons as well as the continuum.

Let us start by recalling that,
in GaAs, the band-edge states at the $\Gamma$ point have a well-defined total angular momentum~\cite{Haug_BookSemiconductors}. The conduction band minimum is twofold degenerate and formed by electron states of angular momentum $J=1/2$ and projections $J_e=\pm 1/2$. Instead, the valence band has a fourfold degenerate structure consisting of $J=3/2$ states with projections $J_h=\pm 1/2, \pm 3/2$. These valence band states correspond to the so-called heavy-hole ($J_h=\pm\tfrac{3}{2}$) and light-hole ($J_h=\pm\tfrac{1}{2}$) states, reflecting their different effective masses near the $\Gamma$ point. In bulk GaAs, the heavy- and light-hole bands are degenerate at the $\Gamma$ point, however this degeneracy is lifted in QWs due to the confinement. 
Within the envelope-function approximation, the exciton wave function near $\Gamma$ is expanded in terms of lattice-periodic Bloch functions $u_{cJ_e} (\r_e)$ and $u_{hJ_h} (\r_h)$ and slowly varying envelope functions $\psi_{J_h}^{J_e}$~\cite{ivchenko2005optical}:
\begin{equation}
\label{eq:X-state-def}
    \Psi^{J_e}(\r_e,\r_h)=\sum_{J_h } \psi_{J_h}^{J_e}(\r_e,\r_h)u_{c J_e}(\r_e)u_{h J_h}(\r_h)\, ,
\end{equation} 
where $\r_{e,h}$ are the three-dimensional coordinates of the electron and hole, which we decompose into their in-  and out-of-plane components, $\r = (\vegr{\rho},z)$. Because we neglect the electron-hole exchange interaction, $J_e$ is a good quantum number; hence we treat the $+1/2$ and $-1/2$ conduction electron sectors separately. Thus, the exciton is described
by a four-component envelope spinor with fixed $J_e$:
\begin{equation}
\label{eq:spinor-envelopes}
    \psi^{J_e}(\r_e,\r_h)=\begin{pmatrix}
        \psi_{\frac{3}{2}}^{J_e}(\r_e,\r_h)\\[5pt]
        \psi_{\frac{1}{2}}^{J_e}(\r_e,\r_h)\\[5pt]
        \psi_{-\frac{1}{2}}^{J_e}(\r_e,\r_h)\\[5pt]
        \psi_{-\frac{3}{2}}^{J_e}(\r_e,\r_h)\\[5pt]
    \end{pmatrix}\, .
\end{equation}
Within the Luttinger Hamiltonian formalism, 
the envelope exciton functions obey the \sch equation:
\begin{equation}
\label{eq:sch-eq-Xs}
    \mathcal{H} \psi^{J_e} (\r_e,\r_h) = E\psi^{J_e}(\r_e,\r_h) \, .
\end{equation}
The Hamiltonian $\mathcal{H}$ has three separate contributions:
\begin{equation}
\label{eq:general-Ham} 
    \mathcal{H} = \mathcal{H}_e + \mathcal{H}_h +\mathcal{V} (|\r_e-\r_h|)  \, .
\end{equation} 
We now go through each of these separately.

The effective-mass Hamiltonian for the conduction electrons $\mathcal{H}_e$
\begin{equation}
\label{eq:e-eff-mass-H}
    {\mathcal{H}}_e = \left[ E_g + \frac{{\kv}_e^2}{2m_e}  +V_e(z_e) +\mu_B g_e J_eB \right] 
    \mathcal{I}_{4}\, ,
\end{equation}
is diagonal in the $4$-dimensional spinor space ($\mathcal I_4$ is the $4\times4$ identity operator). Here, $E_g$ is the bulk GaAs energy gap at the $\Gamma$ point and $m_e$ is the conduction electron effective mass. In the presence of a magnetic field in the $z$-direction, $\vect{B} = (0,0,B)$, the momentum operator $\kv_e$ is (throughout we set $\hbar=1$):
\begin{equation}
    {\kv}_{e} = -i \nabla_{\r_e} + \frac{e}{c} \vect{A}(\r_e)\, ,
\end{equation}
where, in the symmetric gauge, $\vect{A}(\r) = \frac{1}{2}\vect{B} \times \r$. 
Additionally, $V_e(z)= V_c\,\Theta(|z|-d_z/2)$ is the electron confining potential, with $\Theta(x)$ the Heaviside function, $d_z$ the QW width, and $V_{c}$ the conduction band offset. 
Finally, $g_e$ is the electron $g$-factor, and $\mu_B$ the Bohr magneton. 

The second term in Eq.~\eqref{eq:general-Ham}, ${\mathcal{H}}_h$, is the 
Luttinger Hamiltonian  describing valence band holes~\cite{Luttinger_PR1956,ivchenko2005optical}:
\begin{align}
\label{eq:h-eff-mass-H}
    \mathcal{H}_{h}^{} &=  \begin{pmatrix}
       {h}_{\frac{3}{2}} & {b} & {c} & 0\\[5pt]
       {b}^{\,\dag} & {h}_{\frac{1}{2}} & 0 & {c}\\[5pt]
       {c}^{\,\dag} & 0 & {h}_{-\frac{1}{2}} & -{b}\\[5pt]
       0 & {c}^{\,\dag} & -{b}^{\,\dag} & {h}_{-\frac{3}{2}}
   \end{pmatrix} \, ,
\end{align}
where
\begin{subequations}
\begin{align}
    {h}_{\pm\frac{3}{2}} &= \frac{{k}_{z,h}^2}{2m_{z,hh}} + \frac{{k}_{x,h}^2+{k}_{y,h}^2}{2m_{hh}} + V_h (z)\mp 3\mu_B \kappa B\\
    {h}_{\pm\frac{1}{2}} &= \frac{{k}_{z,h}^2}{2m_{z,lh}} + \frac{{k}_{h,x}^2+{k}_{h,y}^2}{2m_{lh}} + V_h (z) \mp \mu_B \kappa B\\
\label{eq:b-term}
    {b} &= -\sqrt{3}\frac{\gamma_3}{m_0}\left\{{k}_{h,z},{k}_{h,x} -i{k}_{h,y}\right\}\\ 
\label{eq:c-term}
    {c} &= -\frac{\sqrt{3}}{2 m_0}\left[\gamma_2 \left({k}_{h,x}^2-{k}_{h,y}^2\right) - 2i \gamma_3 \left\{{k}_{h,x} , {k}_{h,y}\right\}\right]\, .
\end{align}
\end{subequations}
Here, $\gamma_{i=1,2,3}$, and $\kappa$ are the Luttinger parameters~\cite{Luttinger_PR1956}. The $z$-direction and in-plane effective masses for heavy- and light-hole bands are expressed in terms of the Luttinger parameters $\gamma_{1,2}$ as
\begin{align}
\label{eq:mass-and-gammas}
    m_{z,hh,lh} &= \frac{m_0}{\gamma_1 \mp 2 \gamma_2} & 
    m_{hh,lh} &= \frac{m_0}{\gamma_1 \pm \gamma_2} \, ,
\end{align}
where $m_0$ is the free electron mass.  The kinematical momentum for holes is given by
\begin{equation}
    {\kv}_{h} = -i \nabla_{\r_h} - \frac{e}{c} \vect{A}(\r_h)\, .
\end{equation}
As for the electrons, the external confinement takes the form $V_{h} (z) = V_{v}\,\Theta(|z|-d_z/2)$ with $V_v$ the valence band offset.

The terms~\eqref{eq:b-term} and~\eqref{eq:c-term} yield the off-diagonal parts of the Luttinger Hamiltonian. They depend on the anticommutator of momentum operators, where $\{ {a}, {b}\} =  a b +  b a$.
Note that $\gamma_2 \ne \gamma_3$ implies a  broken in-plane rotational symmetry, introduced by the cubic symmetry of the GaAs crystal. However, we can restore the in-plane isotropy~\cite{BastardBook,Bauer_AndoPRB1988} by approximating $\gamma_2 \simeq \gamma_3 \simeq (\gamma_2 + \gamma_3)/2$ in the $c$ term~\eqref{eq:c-term}:
\begin{equation}
\label{eq:c-symmetric}
    {c} \simeq -\frac{\sqrt{3}}{2 m_0}\frac{\gamma_2 + \gamma_3}{2} \left( {k}_{h,x}- i {k}_{h,y}\right)^2 \; .
\end{equation}
This commonly used approximation considerably simplifies the numerical treatment of the exciton problem by effectively reducing the dimension of the Hilbert space, as discussed in the following.

The last term $\mathcal{V} (|\r_e-\r_h|)$ in Eq.~\eqref{eq:general-Ham} describes the electron-hole Coulomb interaction:
\begin{equation}
\label{eq:Coulomb-real}
    \mathcal{V} (|\r_e-\r_h|)=-\frac{e^2}{\varepsilon|\r_e-\r_h|} \mathcal{I}_{4} \, ,
\end{equation}
where $\varepsilon$ is the static dielectric constant, and we employ Gaussian units, $4\pi\epsilon_0 = 1$. Note that we neglect effects due to the small difference in the
dielectric constants of well and barrier materials. 
In GaAs quantum wells with $d_z \gtrsim 10$ nm, this dielectric-constant mismatch introduces corrections smaller than $\Delta E \lesssim 0.1$~meV to the ground-state exciton binding energy~\cite{Shuvayev_2006}\footnote{Note also that we have neglected the spatial variation of the Luttinger parameters and the conduction electron mass across the well–barrier interface, i.e., we model the barrier solely through a confining potential.}.

In a QW, the confining potentials $V_{e,h}(z)$ break translational invariance along the growth direction $z$, so that the $z$ component of the total exciton momentum, $k_{e,z}+k_{h,z}$, is not conserved. By contrast, for a homogeneous magnetic field along $z$, the in-plane component $\mathbf{K}=(k_{e,x}+k_{h,x},k_{e,y}+k_{h,y})$ remains a good quantum number. This quantity, also known as the magnetic momentum, generates translations in the $\vegr{\rho}=(x,y)$ plane. It can be expressed in  terms of the in-plane relative and center of mass coordinates, 
\begin{align}
    \vegr{\rho} &= \vegr{\rho}_e - \vegr{\rho}_h\\
    \vect{R}_{J_h} &= \frac{m_e \vegr{\rho}_e + m_{J_h} \vegr{\rho}_h}{m_e + m_{J_h}}\, ,  
\end{align}
where $m_{\pm 3/2} \equiv m_{hh}$ and $m_{\pm 1/2} \equiv m_{lh}$, as
\begin{equation}
    \mathbf{K}=-i\nabla_{\vect{R}_{J_h}} -\frac{e}{2c}\mathbf{B}\times\rhov \, .
\end{equation}
It is therefore natural to consider the following Lamb-Gor'kov-Dzyaloshinskii ansatz~\cite{Gorkov_SovietJourn1968,Lamb_PR1952} for the 
exciton envelope wave function components~\eqref{eq:spinor-envelopes}: 
\begin{equation}
\label{eq:Lamb-Trans}
    \psi_{J_h}^{J_e} (\r_e,\r_h) = e^{i[\mathbf{K}+\frac{e}{2c}(\vect{B} \times \vegr{\rho})]\cdot \vect{R}_{J_h}} \varphi_{J_h}^{J_e}(\rhov,z_e,z_h)\, .
\end{equation}
Since $\vect{K}$ is a good quantum number, we set it to zero in the following. This corresponds to electron–hole pairs created by photons at normal incidence.
Note that by setting $\mathbf{K}=\0$, the transformation above defines a common reference frame for all the $J_h$ components since the phase $e^{i\frac{e}{2c}(\mathbf{B}\times\rhov)\cdot \vect{R}_{J_h}}=e^{i\frac{eB}{2c}(x_ey_h-y_ex_h)}$  becomes independent of the hole-mass $m_{hh,lh}$.

By substituting the ansatz~\eqref{eq:Lamb-Trans} into Eq.~\eqref{eq:sch-eq-Xs}, one obtains the coupled real-space equations for the components $\varphi_{J_h}^{J_e}(\rhov,z_e,z_h)$, identical to those already reported in Ref.~\cite{Bauer_AndoPRB1988}. Here, however, for numerical convenience we solve the problem in reciprocal space, considering the in-plane Fourier transform from $\vegr{\rho}$ to $\kv$ (we set the system area to $\mathcal{A}=1$):
\begin{equation}
\label{eq:momentum-space-rel-wf}
    \varphi_{J_h}^{J_e}(\rhov,z_e,z_h) = \sum_\kv e^{i\kv \cdot \rhov} \varphi_{J_h \kv}^{J_e}(z_e,z_h)\, .
\end{equation}
Due to the in-plane isotropy of the problem (see Eq.~\eqref{eq:c-symmetric} and surrounding discussion), 
it is advantageous to work in polar coordinates, where 
$\kv=(k,\theta)$ and $\sum_{\kv}\equiv \int_0^{2\pi} d\theta \int_0^{\infty}  dk \,k/(2\pi)^2$.

In in-plane momentum space, the \sch equation~\eqref{eq:sch-eq-Xs} takes the form:
\begin{multline}
\label{eq:X-Luttinger-momentum}
    \tilde{\mathcal{H}} \varphi_{\kv}^{J_e}(z_e,z_h) + \sum_{\kv'} \mathcal{V}_{\kv-\kv'}(z_e-z_h)  \varphi_{\kv'}^{J_e}(z_e,z_h)\\ = E \varphi_{\kv}^{J_e}(z_e,z_h)\, ,
\end{multline}
where $\varphi_{\kv}^{J_e}$ is the $4$-component spinor  $(\varphi_{\frac{3}{2} \kv}^{J_e}, \varphi_{\frac{1}{2} \kv}^{J_e}, \varphi_{-\frac{1}{2} \kv}^{J_e}, \varphi_{-\frac{3}{2} \kv}^{J_e})^T$ and
\begin{equation}
\label{eq:Coulomb-momentum}
    \mathcal{V}_{\kv}(z) = -\frac{2\pi e^2}{\varepsilon k}e^{-|z|k}\,  
    \mathcal{I}_4 \, ,
\end{equation}
is the in-plane Fourier transform of the Coulomb interaction~\eqref{eq:Coulomb-real}. 
In momentum space, the exciton effective-mass Hamiltonian takes the form
\begin{equation}
\label{eq:X-hamil-Lut}
    \tilde{\mathcal{H}} = \begin{pmatrix} 
    {h}_{X\frac{3}{2}} & {b}_{X} & {c}_{X} & 0 \\[5pt] {b}_{X}^{\dagger} & {h}_{X\frac{1}{2}} & 0 & {c}_{X} \\[5pt] {c}_{X}^{\dagger} & 0 & {h}_{X-\frac{1}{2}} & -{b}_{X} \\[5pt] 0 & {c}_{X}^{\dagger} & -{b}_{X}^{\dagger} & {h}_{X-\frac{3}{2}} \end{pmatrix}\, ,
\end{equation}

%
\begin{widetext}
where
\begin{subequations}
\label{eq:expressions-h-b-c}
\begin{align}
    {h}_{X\pm\frac{3}{2}}&=-\frac{\partial_{z_h}^2}{2m_{hh,z}}-\frac{\partial_{z_e}^2}{2m_e} + V_e(z_e) + V_h(z_h) + \frac{k^2}{2\mu_{hh} } 
    - \frac{\mu_{hh}  \omega_{c, hh}^2}{2} \left(\partial_k^2 + \frac{1}{k} \partial_k + \frac{\partial_\theta^2}{k^2}\right) -\omega_{c,hh}\eta_{hh} i\partial_\theta\mp3\kappa\mu_B B \nonumber \\
    &+J_e\mu_B B\\
    {h}_{X\pm\frac{1}{2}}&=  -\frac{\partial_{z_h}^2}{2m_{lh,z}}-\frac{\partial_{z_e}^2}{2m_e} + V_e(z_e) + V_h(z_h) + \frac{k^2}{2\mu_{lh} } 
    - \frac{\mu_{lh}  \omega_{c, lh}^2}{2} \left(\partial_k^2 + \frac{1}{k} \partial_k + \frac{\partial_\theta^2}{k^2}\right) -\omega_{c,lh}\eta_{lh} i\partial_\theta\mp\kappa\mu_B B \nonumber\\
    &+J_e\mu_B B\\ 
   \label{eq:b-coupling-lutt-k}  {b}_{X}&=i\frac{\sqrt{3}\gamma_{3}}{m_0}\partial_{zh} {k}^{-}\\
\label{eq:c-coupling-lutt-k}    
    {c}_{X}&= - \frac{\sqrt{3}}{2m_0} \frac{(\gamma_2+\gamma_3)}{2}({k}^{-})^2\, ,
\end{align}
\end{subequations}
and where we have introduced the following notation for the $k^{\pm}$ operators:
\begin{subequations}
\begin{align}
    {k}^{\pm}=& e^{\pm i\theta}\left[-k+\frac{eB}{2c}\left(\mp\partial_k-i\frac{\partial_\theta}{k}\right)\right]\\
    ({k}^{\pm})^2=& e^{\pm 2i\theta} \left[k^2 - \frac{eB}{2c} \left(\mp2k \partial_k -2i \partial_\theta\right) + \left(\frac{eB}{2c}\right)^2 \left(\partial_k^2 \mp \frac{2i}{k^2} \partial_\theta \pm\frac{2i}{k} \partial_{k\theta} - \frac{\partial_k}{k} - \frac{\partial_\theta^2}{k^2}\right)\right]\, .
\end{align}
\end{subequations}
\end{widetext}
%
Note that the expressions of ${b}^{\dagger}_{X}$ and ${c}^{\dagger}_{X} $ are immediate to derive, since $({k}^{-})^{\dagger}={k}^{+}$. The exciton $hh$ and $lh$ reduced masses $\mu_\sigma$ and mass differences $\eta_\sigma$ are defined as:
\begin{align}
\label{eq:reduced}
    \mu_{\sigma} &= \frac{m_e m_\sigma}{m_e + m_\sigma} & \eta_{\sigma} & =\frac{m_e-m_\sigma}{m_e+m_\sigma}\, ,
\end{align}
with $\sigma=hh,lh$, while the exciton cyclotron frequencies are
\begin{equation}
    \omega_{c,\sigma}=\frac{eB}{2\mu_{\sigma}}\, .
\end{equation}

It is instructive to analyze the symmetries of the eigenvalue problem~\eqref{eq:X-Luttinger-momentum}, as they can be exploited to further simplify the numerical solution. 
As discussed above, the in-plane isotropy of the problem has been restored by approximating $\gamma_2 \simeq \gamma_3 \simeq (\gamma_2 + \gamma_3)/2$  in Eqs.~\eqref{eq:c-symmetric} and~\eqref{eq:c-coupling-lutt-k}. 
Consequently, although the $z$ component of the orbital angular momentum operator of the electron-hole pair, $L_z = -i\partial_\theta$, does not commute with the operators $b_X$ and $c_X$ (and is therefore not conserved), the $z$ component of the exciton total angular momentum operator $\mathcal{J}_z$ is conserved and thus its eigenvalue $J_z$ remains a good quantum number.
The total angular momentum operator $\mathcal{J}_z$ is defined as the following $4\times4$ matrix:
\begin{equation}
    \mathcal{J}_z= \mathcal{J}_h + \left(J_e + L_z\right)
    \mathcal{I}_4\, ,
\end{equation}
where $\mathcal{J}_h = \text{diag}(\frac{3}{2},\frac{1}{2},-\frac{1}{2},-\frac{3}{2})$. 
It is easy to show that 
$\mathcal{J}_z$ commutes with the total Hamiltonian in Eq.~\eqref{eq:X-hamil-Lut}. This follows from
\begin{equation*}
    [(J_e+L_z) \mathcal{I}_4 , \tilde{\mathcal{H}}] = -[\mathcal{J}_h,\tilde{\mathcal{H}}]\,,
\end{equation*}
which can be easily demonstrated by using the identities $[-i\partial_\theta,k^{\pm}] =\pm k^{\pm}$ and $
    [-i\partial_\theta,(k^{\pm})^2]=\pm 2 (k^{\pm})^2$.
For dipole-allowed transitions, the angular momentum $J_z$ is restricted to the values $\pm1$, where $+$ ($-$) corresponds to right- (left-) circularly polarized light.

\vspace{6pt}
\begin{table}
\centering
\renewcommand{\arraystretch}{1.15}
\setlength{\tabcolsep}{6pt}
\begin{tabular}{cc|cccc c|c}
\hline
$J_z$ & $J_e$ 
& $\tfrac{3}{2}$ & $\tfrac{1}{2}$ 
& $-\tfrac{1}{2}$ & $-\tfrac{3}{2}$ 
& $=J_h$ & label \\
\hline\hline
 1  & $-\tfrac{1}{2}$ &  0  &  1  &  2  &  3  & \multirow{4}{*}{$=\ell$} & $hh +$ \\
 1  & $+\tfrac{1}{2}$ & -1  &  0  &  1  &  2  & & $lh +$ \\
-1  & $-\tfrac{1}{2}$ & -2  & -1  &  0  &  1  & & $lh -$ \\
-1  & $+\tfrac{1}{2}$ & -3  & -2  & -1  &  0  & & $hh -$ \\
\hline
\end{tabular}
    \caption{Quartets of allowed values of $(J_h,\ell)$ consistent with the conservation of the $z$-component of the total exciton angular momentum~\eqref{eq:Jz-conservation} for a given sector $\alpha=(J_z,J_e)$. The last column indicates the nomenclature used in this paper for each sector.}
\label{tab:ell-Jh-quartets}
\end{table}
The conservation of $J_z$ and $J_e$ reduces the eigenvalue problem~\eqref{eq:X-Luttinger-momentum} to independent 4-dimensional subspaces for each fixed value of $\alpha = (J_z=\pm 1, J_e=\pm 1/2)$~\cite{Bauer_AndoPRB1988}.
We thus label the eigenfunctions by $\alpha$ and require the angular dependence of each component $J_h$ to be an eigenstate of the orbital angular momentum operator $L_z$ with eigenvalue $\ell$:
\begin{equation}
    \varphi^{\alpha}_{J_h \kv}(z_e,z_h) = e^{i\ell \theta} \varphi^{\alpha}_{J_h k \ell}(z_e,z_h) \delta_{\ell , J_z-J_e-J_h}\,.
\end{equation}
The sectors $\alpha$, together with the corresponding allowed $(J_h,\ell)$ values satisfying
\begin{equation}
\label{eq:Jz-conservation}
J_z = J_h + J_e + \ell \, ,
\end{equation}
are summarized in Table~\ref{tab:ell-Jh-quartets}.

The \sch equation satisfied by $\varphi^{\alpha}_{J_h k \ell}(z_e,z_h)$ can be found by 
inserting the ansatz above into Eq.~\eqref{eq:X-Luttinger-momentum} and acting with the operator $\int_0^{2\pi} \frac{d\theta}{2\pi} e^{-i\ell' \theta}$. We obtain a system of four coupled equations for the sector of the exciton Hilbert space with fixed $\alpha$. By calling $\bar{\ell}=J_z-J_e-3/2$, one gets:
%
\begin{widetext}
\begin{subequations}
\label{eq:X-final-eqs}
\begin{align}
    &{h}_{X \frac{3}{2} \bar{\ell}}\, \varphi_{\frac{3}{2} k \bar{\ell}}^{\alpha}(z_e,z_h) + {b}_{X \bar{\ell}+1}\, \varphi_{\frac{1}{2} k \bar{\ell}+1 }^{\alpha}(z_e,z_h) + {c}_{X \bar{\ell}+2}\, \varphi_{-\frac{1}{2} k \bar{\ell}+2 }^{\alpha}(z_e,z_h) - \int\frac{dk'k'}{2\pi} V_{k k' \bar{\ell}}(z_e-z_h) \varphi_{\frac{3}{2} k' \bar{\ell}}^{\alpha}(z_e,z_h) \nonumber\\
    &= E \varphi_{\frac{3}{2} k \bar{\ell}}^{\alpha}(z_e,z_h) \\
    &{b}_{X \bar{\ell}}^{\dagger}\, \varphi_{\frac{3}{2} k \bar{\ell}}^{\alpha}(z_e,z_h) + {h}_{X \frac{1}{2} \bar{\ell}+1} \, \varphi_{\frac{1}{2} k \bar{\ell}+1}^{\alpha}(z_e,z_h) + {c}_{X \bar{\ell}+3}\, \varphi_{-\frac{3}{2} k \bar{\ell}+3}(^{\alpha}z_e,z_h) - \int\frac{dk'k'}{2\pi} V_{k k' \bar{\ell}+1}(z_e-z_h) \varphi_{\frac{1}{2} k' \bar{\ell}+1}^{\alpha}(z_e,z_h) \nonumber\\ 
    &=E \varphi_{\frac{1}{2} k \bar{\ell}+1}(z_e,z_h) \\
    &{c}_{X \bar{\ell}}^{\dagger}\, \varphi_{\frac{3}{2} k \bar{\ell}}^{\alpha}(z_e,z_h) + {h}_{X-\frac{1}{2} \bar{\ell}+2}\, \varphi_{-\frac{1}{2} k \bar{\ell}+2}^{\alpha}(z_e,z_h) - {b}_{X \bar{\ell}+3}\, \varphi_{-\frac{3}{2} k \bar{\ell}+3}^{\alpha}(z_e,z_h) -\int\frac{dk'k'}{2\pi} V_{k k' \bar{\ell}+2}(z_e-z_h)  \varphi_{-\frac{1}{2} k' \bar{\ell}+2}^{\alpha}(z_e,z_h) \nonumber\\
    &=E \varphi_{-\frac{1}{2} k \bar{\ell}+2}^{\alpha}(z_e,z_h)\\
    &{c}_{X \bar{\ell}+1}^{\dagger}\, \varphi_{\frac{1}{2}k \bar{\ell}+1}^{\alpha}(z_e,z_h) - {b}_{X \bar{\ell}+2}^{\dagger}\, \varphi_{-\frac{1}{2} k \bar{\ell}+2}^{\alpha}(z_e,z_h) + {h}_{X-\frac{3}{2} \bar{\ell}+3}\, \varphi_{-\frac{3}{2} k \bar{\ell}+3}^{\alpha}(z_e,z_h) -\int\frac{dk'k'}{2\pi} V_{k k' \bar{\ell}+3}(z_e-z_h) \varphi_{-\frac{3}{2} k' \bar{\ell}+3}^{\alpha}(z_e,z_h) \nonumber\\
    &=E \varphi_{-\frac{3}{2} k \bar{\ell}+3}^{\alpha}(z_e,z_h)\, .
\end{align}
\end{subequations}
\end{widetext}
%
We denote $h_{XJ_h\ell}$, $b_{X\ell}$, and $c_{X\ell}$ as the expressions in Eqs.~\eqref{eq:expressions-h-b-c} with $L_z \mapsto \ell$. Further, the Coulomb potential~\eqref{eq:Coulomb-momentum} decomposed  in the orbital
angular momentum basis is:
\begin{align}
\label{eq:FT-Coulomb-term-lz}
     V_{kk' \ell}(z)= - \frac{2\pi e^2 }{\varepsilon } \int_0^{2\pi} \frac{d\theta}{2\pi}\cos{(\ell\theta)} 
     \frac{e^{-|z||\kv-\kv'|}}{|\kv-\kv'|}\, ,
\end{align}
where $|\kv-\kv'| = \sqrt{k^2 + k'^2 - 2k k' \cos \theta}$. 

The four coupled equations~\eqref{eq:X-final-eqs} are general and describe the four exciton sectors listed in Table~\ref{tab:ell-Jh-quartets}.
Let us consider, for example, the exciton sector $\alpha=(J_z=+1,J_e=-1/2)$, which can be excited by a photon with right circular polarization. According to Table~\ref{tab:ell-Jh-quartets}, the states $(J_h=3/2,\ell=0)$ (i.e., $hh$ $s$-wave), $(J_h=1/2,\ell=1)$ ($lh$ $p$-wave), $(J_h=-1/2,\ell=2)$ ($lh$ $d$-wave), and $(J_h=-3/2,\ell=3)$ ($hh$ $f$-wave) are hybridized by the effects of the QW confinement and by the magnetic field, i.e., the off-diagonal terms $b_{X\ell}$ and $c_{X \ell}$.
To obtain the corresponding energies, one must solve the coupled equations~\eqref{eq:X-final-eqs} with $\bar{\ell}=0$. 
Instead, for, e.g., the sector $\alpha=(J_z=+1,J_e=+1/2)$, one has to impose $\bar{\ell}=-1$. 
A nomenclature note is due: Despite valence-band mixing, in the following we refer to the exciton sector with $\alpha = (J_z = \pm 1, J_e = \mp 1/2)$ as the $hh$ exciton with $\pm$ polarization, and the sector with $\alpha = (J_z = \pm 1, J_e = \pm 1/2)$ as the $lh$ exciton with $\pm$ polarization; this notation is also indicated in Table~\ref{tab:ell-Jh-quartets}.
We numerically solve Eqs.~\eqref{eq:X-final-eqs} in all four sectors, as we discuss below in Sec.~\ref{sec:results}.

Importantly, only the $s$-wave component carries a finite oscillator strength, as the exciton oscillator strength is proportional to the exciton recombination probability~\cite{Yu-CardonaBook} :
\begin{equation}
\label{eq:recom-prob}
    f_\alpha = \left|\int_{-\infty}^{\infty} dz \int \frac{dk k}{2\pi} \varphi^\alpha_{J_h=J_z-J_e k \ell=0} (z,z)\right|^2\, .
\end{equation}
%
%
%

%

\subsection{Model parameters}
\label{sec:model-parameters}
The microscopic model parameters used to determine the exciton energies are summarized in Table~\ref{tab:parameters-model}. These parameters, specific to GaAs/Al$_{0.4}$Ga$_{0.6}$As QWs, are taken from the literature (see table caption).

\begin{table}[h]
\centering
\begin{tabular}{l l c}
\hline
 & \textbf{Parameter} &  \textbf{Value} \\
\hline\hline
electron mass 
 & $m_{e}$ ($m_0$) & $0.067$ \\
dielectric constant
 & $\varepsilon$ & 12.5 \\
electron $g$-factor
 & $g_e$ & -0.44 \\
 \hline
\multirow{4}{*}{Luttinger parameters} 
 & $\gamma_{1}$ & 6.98 \\
 & $\gamma_{2}$ & 2.06 \\
 & $\gamma_{3}$ & 2.93 \\
 & $\kappa$ & 1.2 \\
\hline
\multirow{3}{*}{Band offset} 
 & $\Delta E_g$~(eV)  & 0.5  \\
 & $V_v (\Delta E_g)$ & $0.33$ \\
 & $V_c  (\Delta E_g)$ & $0.67$ \\
\hline
\end{tabular}
\caption{Microscopic model parameters specific to GaAs/Al$_{0.4}$Ga$_{0.6}$As QWs.  The Luttinger parameters have been taken from Ref.~\cite{Vurgaftman-GaAsPar_JAP2001}, while the electron mass and the electron $g$-factor from Ref.~\cite{Bauer_AndoPRB1988}.
At $4$~K, the values of the badgaps of bulk GaAs and Al$_{0.4}$Ga$_{0.6}$As are $E_g^{\text{bulk}}(\text{GaAs})=1.519$~eV and $E_g^{\text{bulk}}(\text{Al$_{0.4}$Ga$_{0.6}$As})\simeq2.055$~eV~\cite{Vurgaftman-GaAsPar_JAP2001}. This gives a bandgap difference
$\Delta E_g\simeq0.5$~eV. The band-gap discontinuity is assumed to be partitioned between the conduction and valence bands in a $67{:}33$ ratio~\cite{Bataev-Ignatiev-Efimov_PRB2022}, such that $V_c + V_v = \Delta E_g$.}
\label{tab:parameters-model}
\end{table}

\section{Results}
\label{sec:results}
%
We now analyze the exciton properties derived from the Luttinger model introduced in the previous section. 
We first summarize the numerical procedure that we employ to solve the coupled exciton Schrödinger equations. We then analyze the exciton properties at zero magnetic field, characterizing the ground and Rydberg excited states  of both $hh$ and $lh$ excitons, as well as the dependence of band hybridization on the QW width. Finally, we study the exciton properties at finite magnetic field for a wide QW of width $d_z=18$~nm, relevant for the experimental setup. Comparison with experiments is presented in Sec.~\ref{sec:experiments}.

\subsection{Numerical implementation} 

The Schrödinger equation for a QW exciton in a magnetic field within the Luttinger formalism has already been solved numerically in both real and momentum space in several previous works~\cite{Bauer_AndoPRB1988,Fancilotto_SLattM1987,Yang_PRL1987,Grigoryev_PRB2016,Grigoryev-Chukeev_JAP2025}. A common approach is to expand the exciton envelope function on a sufficiently large basis and then diagonalize the corresponding Hamiltonian matrix. For example, Ref.~\cite{Bauer_AndoPRB1988} employs the bound-state wave functions of a two-dimensional hydrogen atom for the in-plane degrees of freedom. However, this approach converges poorly at strong magnetic fields. Alternatively, Landau levels can be used as a basis~\cite{Yang_PRL1987}, although they exhibit the opposite limitation. Other more recent approaches take as their starting point the exciton wavefunction eigenstates of the Luttinger Hamiltonian with the off-diagonal terms set to zero~\cite{Grigoryev_PRB2016}. The importance of including the in-plane electron–hole continuum for each confined subband, even at $B=0$, was already emphasized in Refs.~\cite{Andreani_EurPhys1988,Andreani-Pasquarello_PRB1990} as essential for a proper description of the interplay between Coulomb interaction and band mixing.

Here, we treat the in-plane degrees of freedom by exact diagonalization, allowing Coulomb interaction, valence-band mixing, and magnetic-field effects to be included on an equal footing.
This is carried out by discretizing the momentum $k$ on a Gauss-Legendre grid and evaluating the momentum derivatives via a midpoint finite difference scheme---for details see App.~\ref{App:finite-diff}. 
The Coulomb potential~\eqref{eq:FT-Coulomb-term-lz} has an integrable pole for $k=k'$, which we regularize by employing a Land\'e subtraction scheme. This approach has been shown to accelerate numerical convergence in analogous 2D exciton~\cite{Laird_PRB2022,deLaFuentePico_PRB2025} and trion~\cite{Kumar-Levinsen_PRB2025} problems. We demonstrate in App.~\ref{App:subs-scheme} how this scheme can be generalized to the case of finite-width QWs.

Conversely, we use an expansion over a finite basis of confined states for the out-of-plane degrees of freedom $z_e,z_h$~\cite{Bastard_1999JQE,Bauer_AndoPRB1988}:
\begin{equation}
\label{eq:expansion-basis-confined}
    \varphi_{J_h k \ell}^{\alpha}(z_e,z_h)=\sum_{n_e,n_h} \zeta_{n_e}(z_e) \xi_{J_h n_h}(z_h) \varphi^{\alpha}_{J_h k \ell n_e n_h}\, ,
\end{equation}
where $\zeta_{n_e} (z)$ and $\xi_{J_h n_h}(z)$ are the solutions of the uncoupled 1D \sch equations describing the electron and hole transverse motion: 
%
\begin{subequations}
\label{eq:transverse}
\begin{align}
    \left[-\frac{\partial_z^2}{2m_e} + V_{e}(z)\right] \zeta_{n_e}(z) =& \epsilon_{n_e} \zeta_{n_e}(z) \\
    \left[-\frac{\partial_z^2}{2m_{J_h}} + V_{h}(z)\right] \xi_{J_h n_h}(z) =& \varepsilon_{J_h n_h}\xi_{J_h n_h} (z)\, .
\end{align}
\end{subequations}
Naturally, $\zeta_{n_e}(z)$ and $\xi_{J_h n_h} (z)$ form an orthonormal basis with a well defined parity due to the $z \to -z$ symmetry of the well---i.e., $\zeta_{n} (-z) = (-1)^{n} \zeta_{n}(z)$ and $\xi_{J_h n} (-z) = (-1)^n \xi_{J_h n} (z)$ ($n \in \mathbb{N}$). 
Thus, the off-diagonal term $b_X$ in the exciton effective-mass Hamiltonian~\eqref{eq:X-hamil-Lut}, being proportional to $\partial_{z_h}$~\eqref{eq:b-coupling-lutt-k}, mixes $hh$ and $lh$ states belonging to subbands of opposite parity in the $z_h$ coordinate, whereas the other off-diagonal term $c_X$~\eqref{eq:c-coupling-lutt-k} couples states with the same parity. By contrast, in the $z_e$ coordinate, electron states from different confinement subbands 
are only mixed via the electron-hole Coulomb interaction, which preserves the parity of the electron–hole pair, i.e., it does not change $(-1)^{n_e+n_h}$.
We truncate the number of electron, $hh$, and $lh$ confined states at values sufficient to achieve numerical convergence. Details of the convergence with respect to the number of confined states and the number of $k$-grid points are provided in App.~\ref{app:convergence}.
We now discuss our numerical results for the exciton properties at zero and finite magnetic field.

\begin{figure}
\centering
\includegraphics[width=\linewidth]{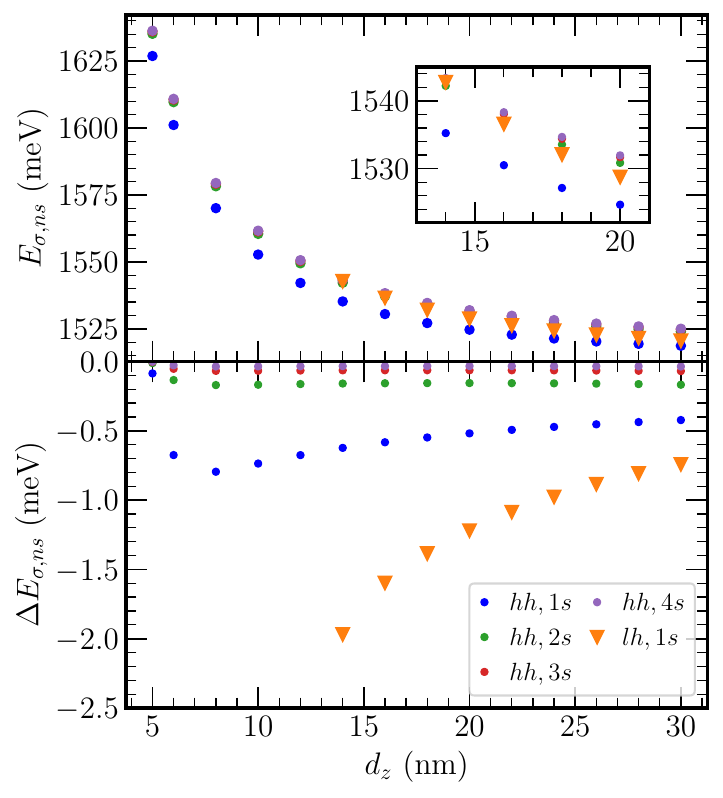}
    \caption{Top panel: Energies $E_{\sigma,ns}$ of the first four $\sigma=hh$ exciton states (circles) and the $\sigma=lh$ ground exciton state (triangles) as a function of QW width at zero magnetic field. Bottom panel: Band-mixing corrections $\Delta E = E - E_{\mathrm{diag}}$ for the same states, where $E_{\mathrm{diag}}$ are exciton energies evaluated by neglecting off-diagonal terms of the Luttinger Hamiltonian~\eqref{eq:X-hamil-Lut}.
    }
\label{fig:Energy-LZdependence}
\end{figure}
%
%
\subsection{Zero magnetic field}
\label{eq:zero-field}
A 2D hydrogenic description of excitons in GaAs QWs breaks down for well widths $d_z \gtrsim 10$~nm due to valence-band mixing and modifications of the Coulomb interaction associated with reduced confinement. In particular, valence-band mixing is known to enhance the binding energies of the $hh,1s$ and $lh,1s$ excitons relative to calculations that neglect the off-diagonal terms of the Luttinger Hamiltonian~\eqref{eq:X-hamil-Lut}~\cite{Andreani_EurPhys1988,Ekenberg-Altarelli_PRB1987,Andreani-Pasquarello_PRB1990,Zhu-HuangPRB1987}. This enhancement is more pronounced for the $lh$ exciton than for the $hh$ exciton, particularly at smaller QW widths, where the light-hole $1s$ exciton approaches the $hh$ continuum~\cite{Andreani_EurPhys1988}.
In this section, for completeness, we briefly summarize the main properties of both $hh$ and $lh$ exciton states at zero magnetic field across different QW widths, providing a basis for the discussion of their behavior in finite magnetic fields in the following section.

The energies of the lowest-lying $hh$ exciton states and the $lh$ exciton ground state as a function of quantum-well width are shown in Fig.~\ref{fig:Energy-LZdependence}, where  
we only plot states that are predominantly composed of the lowest confined state with $n_e = n_h = 0 $.
Despite the valence-band mixing effects discussed above, we label the different $hh$ excitons  
with $ns$, where $n$ has the role of principal quantum number and $s$ stands for $s$-wave symmetry. This is because one can show that these states retain a predominantly $s$-wave–like character, with the associated wave function preserving the expected in-plane nodal structure, where the $ns$ state exhibits $n-1$ nodes (see Section~\ref{sec:results-MF} below). For the $lh$ ground state, we use the same notation, even though for small QW widths (not shown in Fig.~\ref{fig:Energy-LZdependence}) the lowest-energy $lh$ state becomes degenerate with the $hh$ continuum and strongly mixes with it. 
In particular, in agreement with previous works~\cite{Andreani_EurPhys1988,Ekenberg-Altarelli_PRB1987,Andreani-Pasquarello_PRB1990}, our results show that the $lh$ ground-state exciton lies within the electron-hole continuum of the $hh$ sector when $d_z \lesssim 15$~nm.  
A more detailed discussion of the $lh$ state in narrow quantum wells can be found in Refs.~\cite{Andreani_EurPhys1988,Andreani-Pasquarello_PRB1990}.
For increasing well width, the $1s$ $lh$ energy instead approaches that of the $hh$ ground state such that when $d_z \gtrsim 16$~nm the $lh$ ground state falls between the $hh$ ground state and the first excited $hh$ exciton.

Furthermore, Fig.~\ref{fig:Energy-LZdependence} shows the valence-band mixing correction for these states, defined as the difference between the exciton energies obtained from Eqs.~\eqref{eq:X-final-eqs} and those calculated by neglecting the off-diagonal terms of the Luttinger Hamiltonian in Eq.~\eqref{eq:X-hamil-Lut}.
Valence-band mixing reduces the exciton energy, thereby increasing the exciton binding energy~\cite{Andreani-Pasquarello_PRB1990}.
The absolute value of the energy difference shows that the mixing corrections are generally small, remaining below $\sim 1$~meV for the $hh$ ground state, and below $\sim 0.2$~meV for the excited states. They are, however, larger (of the order of $2$~meV) for the $lh$ exciton in narrow quantum wells~\cite{Andreani_EurPhys1988}, due to strong mixing with the electron-hole $hh$ continuum. 
We observe a nonmonotonic behavior of the energy difference for the $hh$ states in narrow quantum wells ($5 \lesssim d_z \lesssim 15$~nm), arising from competing effects~\cite{Bataev-Ignatiev-Efimov_PRB2022}: As the quantum-well width increases, the $lh$ $p$- and $d$-wave states approach the $hh$ $s$-wave states in energy (enhancing mixing), while the mixing matrix elements decrease due to stronger momentum-space confinement of the excitons ($b_X, c_X \propto k, k^2$), which reduces mixing. 
For $d_z \gtrsim 8$~nm, the energy separation between the coupled states varies only weakly with well width; consequently, the reduction of the mixing matrix elements becomes the dominant effect.
Therefore, the mixing corrections to the $hh$ energies first increase and then decrease with increasing $d_z$. 
Finally, for a given quantum-well width, the energy corrections due to valence-band mixing are smaller for excited states. This is expected, since ${b}_{X},{c}_{X} \propto k,k^2$, and higher Rydberg excitons are progressively more confined in momentum space. 
\begin{figure}
\centering
\includegraphics[width=\linewidth]{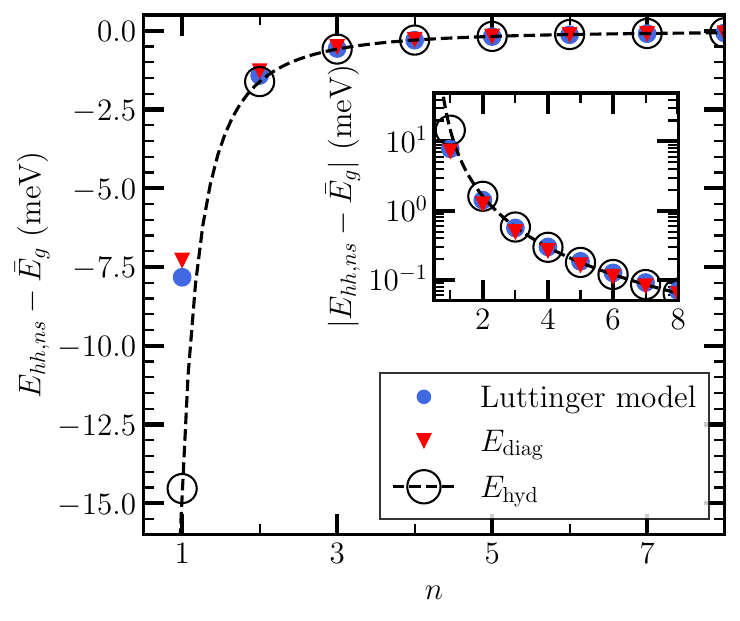}
    \caption{Lowest-lying $hh$ exciton energies at zero magnetic field and $d_z=18$~nm, measured relative to the continuum edge~\eqref{eq:2D-hydrogen-model}. Blue dots are obtained from the full solution of Eq.~\eqref{eq:X-final-eqs}, with the 
    red triangles corresponding to $E_{\text{diag}}$ obtained by neglecting the off-diagonal terms involving 
    $b_X$ and $c_X$. 
    The dashed line and open black circles show the 2D hydrogenic model $E_{\text{hyd}}$~\eqref{eq:2D-hydrogen-model} with $R_{Xhh}=13.1$~meV. The inset shows the same energies on an absolute logarithmic scale, highlighting the agreement between the models for the Rydberg states.}
\label{fig:non-hydrogenic-series}
\end{figure}%

In Fig.~\ref{fig:non-hydrogenic-series}, we compare the $hh$ exciton energies for a fixed width $d_z=18$~nm to a 2D hydrogenic model:
\begin{align}
\label{eq:2D-hydrogen-model}
    E_{\text{hyd}} = \underbrace{E_g + \epsilon_0 + \varepsilon_{\frac{3}{2}0}}_{\displaystyle \bar{E}_g} - \frac{R_{Xhh}}{(2n-1)^2} \, .
\end{align}
We plot only the states that are predominantly composed of the lowest confined state, $n_e = n_h = 0 $, leading to a shifted effective band gap $\bar{E}_g$ due to the zero-point energy in the transverse direction---see Eq.~\eqref{eq:transverse}.
In Eq.~\eqref{eq:2D-hydrogen-model}, $R_{X hh} = 2\mu_{hh} e^4/\varepsilon^2 \simeq 14.53$~meV is the 2D $hh$ exciton Rydberg energy, whose value is obtained by using the parameters of Table~\ref{tab:parameters-model} and the expression~\eqref{eq:reduced} and~\eqref{eq:mass-and-gammas} for the reduced mass.
The full calculation gives instead a binding energy of $7.79$~meV, i.e., considerably smaller, showing that the $hh$ exciton ground state is strongly non-hydrogenic. The main deviation comes from the finite width of the QW, while valence-band mixing contributes only a small correction ($\lesssim 0.5$~meV; see Fig.~\ref{fig:Energy-LZdependence}).
By contrast, excited Rydberg states are better described by a hydrogenic series. Being more extended in real space, they are less sensitive to the reduction in Coulomb attraction due to the finite QW width, since $e^{-|z|k} \simeq 1$ in Eq.~\eqref{eq:Coulomb-momentum}. Similarly, we see that the excited states are well described by keeping only diagonal terms in Eq.~\eqref{eq:X-final-eqs}, as anticipated above.

\subsection{Finite magnetic field}
\label{sec:results-MF}
We now investigate the effects of a static magnetic field applied perpendicular to the QW. 
As we will show, the in-plane localization induced by the magnetic field---together with the modified energy separation between coupled states---leads to a strong enhancement of the mixing between the bright $hh$ $s$-wave states and the dark $lh$ $p$- and $d$-wave states. In contrast to the zero-magnetic-field case, the higher excited Rydberg $hh$ states then become more strongly admixed than the $hh$ ground state.

\begin{figure}
\centering   \includegraphics[width=\linewidth]{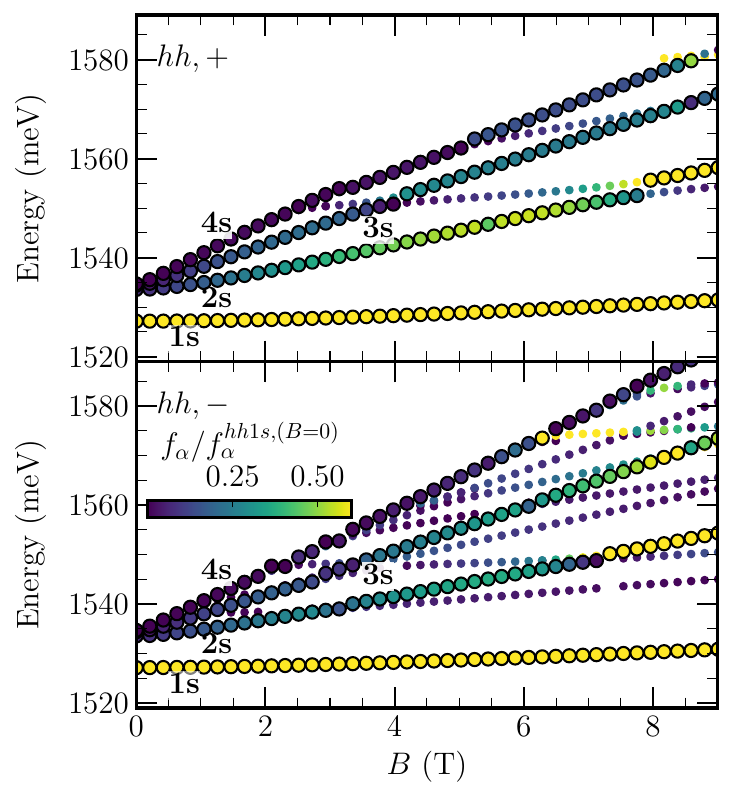}
    \caption{ Exciton energies as a function of magnetic field obtained by solving Eq.~\eqref{eq:X-final-eqs} numerically for the $hh+$ (top) and $hh-$ (bottom) sectors (see Table~\ref{tab:ell-Jh-quartets}). The dot color represents the oscillator strength~\eqref{eq:recom-prob}, rescaled with respect to the value of the ground state at $B=0$. We only show states with oscillator strength $f_{\alpha}/f_{\alpha}^{hh,1s(B=0)} > 0.0016$. Enlarged circled dots highlight the states satisfying the additional condition $w_{00}^{hh\pm} > 0.18$~\eqref{eq:nenh-swave-fraction}.}
\label{fig:Xenergies-colored-osc-str}
\end{figure}

We start from the $hh \pm$ sectors (see Table~\ref{tab:ell-Jh-quartets}) and plot in Fig.~\ref{fig:Xenergies-colored-osc-str} the exciton energies as a function of the magnetic field.
The color of each point represents the oscillator strength~\eqref{eq:recom-prob} and, in order to identify the states that can be effectively measured experimentally, we highlight those whose oscillator strength exceeds a minimal value.
Generally, we see that as the magnetic field increases, anticrossings occur with states that at $B=0$ lie above the $hh$ continuum. These are either $hh$ states of higher confinement but still $1s$ in-plane character (recognizable by their weaker magnetic-field dispersion), or---via valence-band mixing---states of predominantly $lh$ character\footnote{In the energy range considered here, this is expected to occur mainly with $lh, d$ states from the subbands $n_e=0, n_{lh}=0$.}.

Furthermore, we aim to single out exciton states with the same overall symmetry, labeling them according to their dominant $s$-wave and subband character, i.e., their dominant $(n_e,n_h)$ confinement components. 
Thus, we evaluate the $s$-wave and $(n_e,n_h)$ confinement fraction as:
\begin{subequations}
\label{eq:nenh-swave-fraction}
\begin{align}
\label{eq:nenh-hh-swave-fraction}
    w^{hh s \pm}_{n_e n_h} &= \int \frac{dk k}{2\pi}\left|\varphi^{J_z=\pm1 ,J_e=\mp\frac{1}{2}}_{J_h=\pm\frac{3}{2},k, \ell=0, n_e n_h}\right|^{2} \\
\label{eq:nenh-lh-swave-fraction}
    w^{lh s \pm}_{n_e n_h} &=\int \frac{dk k}{2\pi}\left|\varphi^{J_z=\pm 1,J_e=\pm\frac{1}{2}}_{J_h=\pm\frac{1}{2}, k, \ell=0, n_e n_h}\right|^{2} ,
\end{align}    
\end{subequations}
%
and we identify in Fig.~\ref{fig:Xenergies-colored-osc-str} the exciton states with the largest $s$-wave and ground-state confinement fraction $w^{hh s \pm}_{00}$. 
In this way, we isolate the $hh$ states denoted as $ns,\pm$, with $n=1,\dots,4$. However, as noted before, this is only a labeling, since these states exhibit increasing admixture with states of different symmetries, which grows with both energy and magnetic field. In fact, we can only follow the $hh$ $4s$ exciton up to $B=8.5$~T as a consequence of the mixing with a large number of states. 

\begin{figure}
    \centering
    \includegraphics[width=1.0\linewidth]{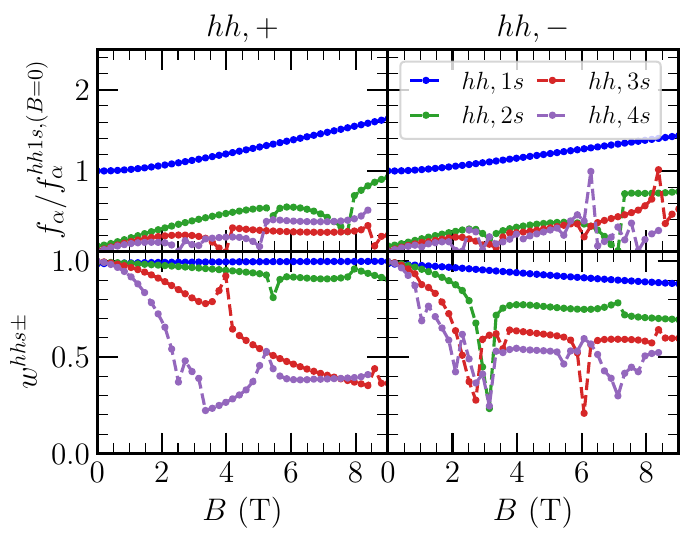}
    \caption{Top: oscillator strength ~\eqref{eq:recom-prob} of the states identified as $hh, ns$ Rydberg states in Fig.~\ref{fig:Xenergies-colored-osc-str}, rescaled with respect to the value of the ground state at $B=0$. Bottom: corresponding
    $s$-wave fraction $w^{hh s}$~\eqref{eq:s-wave-fraction}.}
    \label{fig:ws-RydbergX}
\end{figure}

The color coding in Fig.~\ref{fig:Xenergies-colored-osc-str} highlights that there is a general increase of oscillator strength with increasing magnetic field. 
This trend is further investigated in Fig.~\ref{fig:ws-RydbergX}, which indeed shows
that the oscillator strength of each $hh, ns$ state initially increases with magnetic field up to $B \lesssim 2$~T, due to the in-plane confinement of the exciton wave function induced by the field.  
For larger $B$, we see that whenever an avoided crossing occurs one of the two branches loses oscillator strength as the $s$-wave and ground-state confinement fraction decreases, while the other gains strength. Following the branch with the larger $s$-wave and confinement fraction $w_{00}^{hhs\pm}$ leads to the non-monotonic behavior observed in Fig.~\ref{fig:ws-RydbergX}.

Furthermore, we plot in the lower panels of Fig.~\ref{fig:ws-RydbergX} the total $s$-wave fraction for each of these states, defined as:
\begin{equation}
\label{eq:s-wave-fraction}
w^{(hh,lh)s\pm} = \sum_{n_e,n_h} w^{(hh,lh)s\pm}_{n_e n_h} \, .
\end{equation}
We observe that the $1s$ state remains predominantly $s$-wave, although in the $-$ polarization the $s$-wave fraction shows a slight monotonic decrease.
For the excited states, the $s$-wave fraction decreases more rapidly with magnetic field, exhibiting resonant-like dips or peaks associated with anticrossings with other states: Dips occur when the coupled state has a smaller $s$-wave fraction (such as $lh,d$ states from the $(n_e=0,n_{lh}=0)$ subbands), while peaks occur when it is larger (such as $hh,1s,\pm$ states from higher confinement subbands of the same electron–hole parity).
When the $s$-wave fraction drops below 50\%, as in the case of the $3s$ and $4s$ states already at moderate magnetic fields, we retain the same notation even though the degree of admixture with other states becomes significant. 
We analyze in App.~\ref{app:Rydberg-comp} the contributions of the higher-$\ell$ components to the $hh$ states, i.e., the $p$-, $d$-, and $f$-wave fractions.

\begin{figure}
    \centering
    \includegraphics[width=1.0\linewidth]{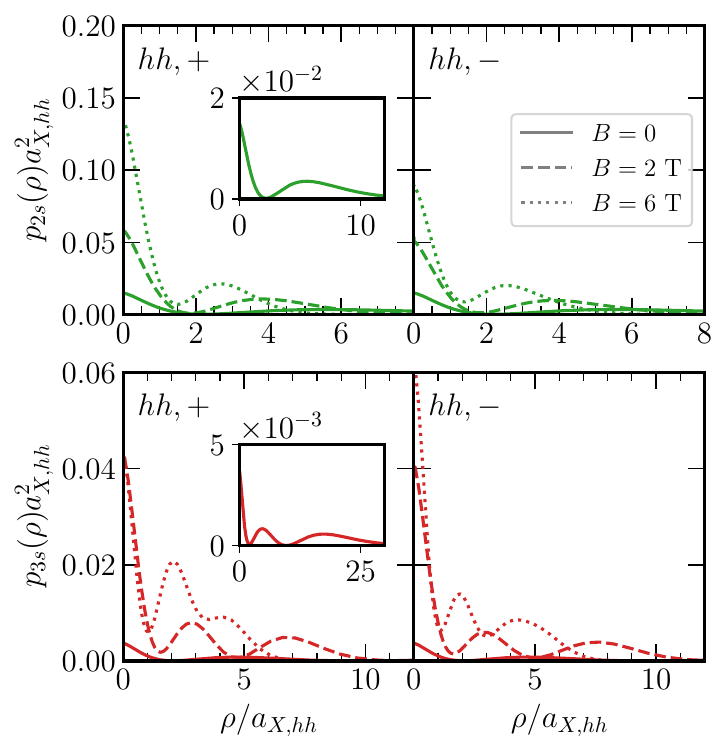}
    \caption{In-plane probability density~\eqref{eq:prob-density} for the $hh$ exciton states labeled as $2s$ (top) and $3s$ (bottom) at different values of the magnetic field. Here, $a_{X\sigma}=\varepsilon/(2\mu_\sigma e^2)$ is the 2D exciton Bohr radius. The inset shows the $B=0$ case with a reduced $y$-axis range. }
\label{fig:in-plane-prob-density}
\end{figure}

This behavior is also reflected in the nodal structure of the in-plane exciton probability density $p(\rho)$, which is defined as\footnote{Note that $p(\rho)$ can be obtained starting from the full exciton probability density $|\Psi^{J_e}(\mathbf{r}_e,\mathbf{r}_h)|^2$ in Eq.~\eqref{eq:X-state-def}, including the lattice-periodic Bloch functions $u_{c,v}$. The orthogonality of the hole Bloch functions with different $J_h$ leads to the same expression as in Eq.~\eqref{eq:prob-density}, provided that variations of the exciton wave function on the length scale of the lattice are neglected.}:
\begin{equation}
\label{eq:prob-density}
    p(\rho) = \sum_{J_h} \int dz_e dz_h |\varphi^{\alpha}_{J_h}(\rhov,z_e,z_h)|^2\, .
\end{equation}
In Fig.~\ref{fig:in-plane-prob-density}, we plot $p(\rho)$ for two $hh$ excited states at different magnetic fields. We observe that the in-plane nodal structure, with the $ns$ state exhibiting $n-1$ nodes as expected for the ideal 2D case~\cite{Laird_PRB2022}, is preserved, although at higher magnetic fields the nodes evolve into minima rather than strict zeros. 

\begin{figure}
    \centering
    \includegraphics[width=\linewidth]{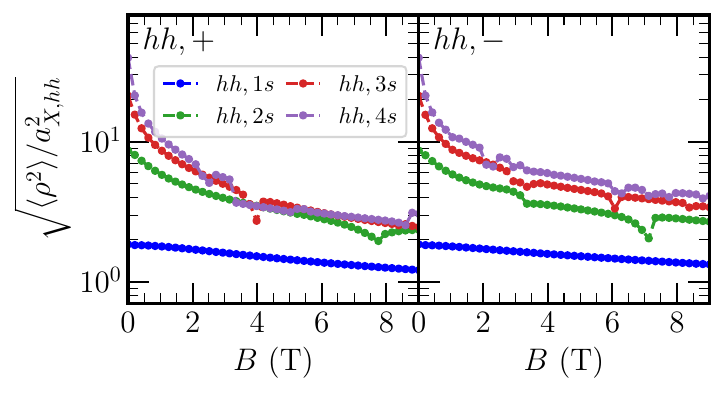}
    \caption{Root-mean-square electron-hole in-plane distance~\eqref{eq:in-plane-rel-dist} of the exciton states labeled as $hh, ns$ ($n=1, \dots, 4$) as a function of magnetic field.}
    \label{fig:electron-hole-distance-Rydberg}
\end{figure}

Similarly, we can study the average in-plane electron–hole relative distance of the exciton states and how it is affected by band mixing. We define:
\begin{equation}
\label{eq:in-plane-rel-dist}
   \langle \rho^2\rangle = \sum_{J_h}\int d\rhov dz_e dz_h \rho^2|\varphi_{J_h}^{\alpha}(\rhov,z_e,z_h)|^2 \, ,
\end{equation}
and plot the magnetic field dependence in Fig.~\ref{fig:electron-hole-distance-Rydberg} for the $hh$ states. Overall, we recover the expected reduction of the in-plane electron–hole separation with increasing magnetic field. In addition, band mixing further enhances the confinement of the exciton wave function. For instance, the $3s,+$ state is more strongly mixed than the $2s,+$ state, and, as a result, the two acquire similar radii for $B \gtrsim 4$~T. By contrast, in the $-$ sector, the $s$-wave fraction is similar, and the radii of the $3s$ and $2s$ states are clearly distinct.

\begin{figure}
    \centering
    \includegraphics[width=\linewidth]{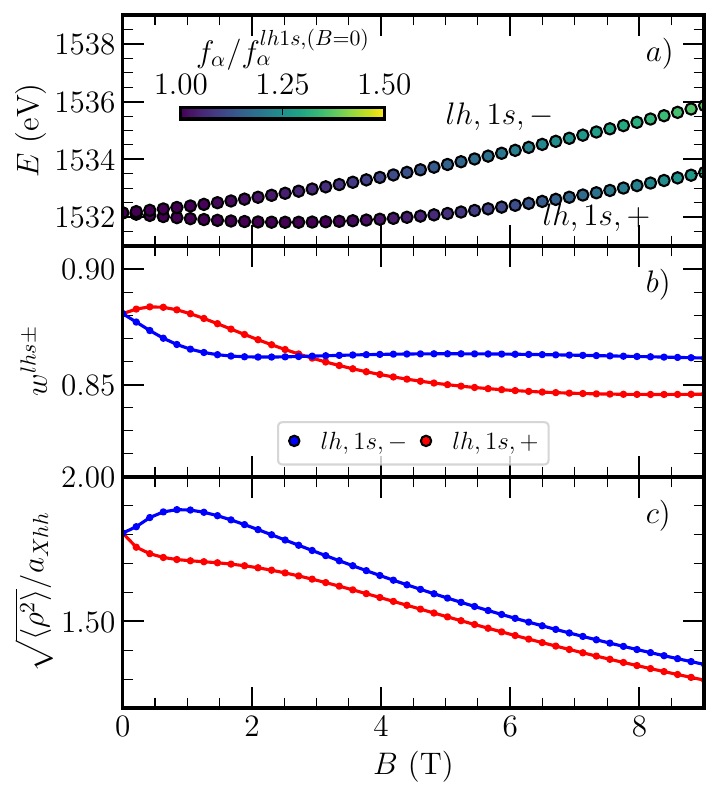}
    \caption{ (a) Energies of the $lh,1s,\pm$ excitons as a function of magnetic field. The color scale indicates the oscillator strength $f_\alpha$, normalized to its value at $B=0$, $f_{\alpha,lh1s}^{B=0}$. (b) Corresponding light-hole $s$-wave fraction $w^{lh s}$, defined in Eq.~\eqref{eq:s-wave-fraction}. (c) In-plane root-mean-square electron-hole distance, defined in Eq.~\eqref{eq:in-plane-rel-dist}.}
\label{fig:lh-1s-theory}
\end{figure}

We finally turn to the $lh,\pm$ sector, focusing on the ground state only. This is because the excited states lie above the electron–hole continuum of the $hh,\pm$ sector, where strong intersubband mixing makes it difficult to isolate the $lh$ excited states. Moreover, these states are not relevant for the comparison with the experimental results in the following section.
We plot in Fig.~\ref{fig:lh-1s-theory} the energy, $s$-wave fraction, and electron–hole relative distance for the $lh,\pm$ states. We observe a monotonic increase of the oscillator strength with increasing magnetic field. It is interesting to note that the radius of the $-$ state has a nonmonotonic behavior with the magnetic field. At small $B$, it increases due to enhanced mixing with $hh,d,-$ states, as reflected by the decrease of the $s$-wave fraction $w^{lh s}$. At larger magnetic fields, $w^{lh s}$ saturates and the electron–hole average distance decreases due to magnetic confinement. By contrast, the radius of the $+$ state decreases monotonically, with a stronger reduction at small magnetic fields driven by the initial increase of $w^{lh s}$.

\begin{figure}
    \centering
    \includegraphics[width=\linewidth]{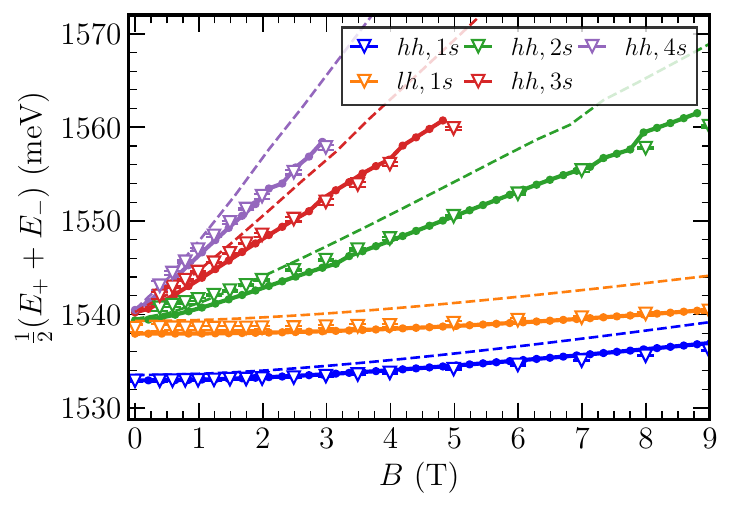}
    \caption{Comparison between the experimental (triangles) and theoretical (solid lines) exciton diamagnetic shift as a function of the  magnetic field. The dashed lines are evaluated by setting the off-diagonal terms of the Luttinger Hamiltonian to zero.}
\label{fig:diamagnetic-shift}
\end{figure}
\begin{figure*}
    \centering
  \includegraphics[width=\linewidth]{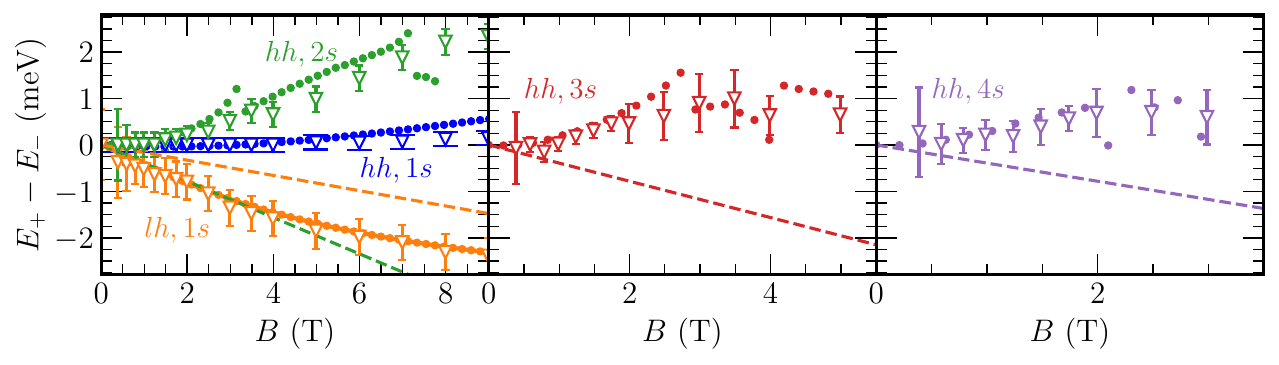}
    \caption{Comparison between the experimental (triangles) and theoretical (dots) exciton Zeeman splitting as a function of the magnetic field. The dashed lines are evaluated by setting the off-diagonal terms of the Luttinger Hamiltonian to zero.}
\label{fig:zeeman}
\end{figure*}
%
\section{Comparison to experiment}
\label{sec:experiments}
We compare the theoretical results with polarization-resolved reflectance measurements performed on a heterostructure sample incorporating 12 GaAs/Al$_{0.4}$Ga$_{0.6}$As QWs. The sample slab is described in detail in the accompanying paper~\cite{short-paper}; in that work, 
a portion of the slab was patterned with a one-dimensional photonic crystal grating. This patterning enabled the investigation of the very strong coupling regime between excitons and the slab waveguide modes exposed to an applied magnetic field. Crucially, that regime featured  
an additional and much stronger hybridization between $hh$ and $lh$ exciton states, now mediated by their coupling to light.
By contrast, the measurements presented here are performed on an unpatterned region of the sample, resulting in the absence of photonic modes within the investigated energy range. The system therefore remains in the weak coupling regime, allowing us to focus entirely on the excitonic properties.

The sample is mounted in a closed-cycle cryostat at $4$~K equipped with a superconducting magnet, which provides magnetic fields up to $B=9$~T in Faraday geometry. The sample is illuminated with a thermal white light source with the reflected light passing through a circular polarizer to distinguish the two polarization states. We adopt the convention introduced in Sec.~\ref{sec:model}, where $\pm$ circularly polarized light corresponds to $J_z=\pm1$. Further details of the experimental procedure and the extraction of the exciton energies can be found in the accompanying paper, Ref.~\cite{short-paper}. 

From the polarization-resolved energies, we compare the theoretical and experimental result 
for the diamagnetic shift and the Zeeman splitting in Figs.~\ref{fig:diamagnetic-shift} and~\ref{fig:zeeman}, respectively. The error bars in these figures represent estimates on the uncertainty of the exciton energy extraction procedure. These uncertainties arise primarily from background noise at low magnetic fields, which particularly affects the excited states due to their low oscillator strength, and from the spectral overlap with additional states at high magnetic fields (cf.~Fig.~\ref{fig:Xenergies-colored-osc-str}).
For the theoretical results we consider only the exciton energies highlighted in Figs.~\ref{fig:Xenergies-colored-osc-str} and~\ref{fig:lh-1s-theory}, which have above-threshold values of oscillator strength and $s$-wave fraction.  
Similarly to the theoretical results, the experimental reflectance spectra exhibit a proliferation of excited states as the magnetic field increases. As a consequence, an unambiguous identification of the $hh,3s$ and $hh,4s$ states beyond $B=4$~T becomes increasingly challenging. Although these states could, in principle, be assigned based on their agreement with theory, we avoid such a procedure and restrict our analysis to $B \lesssim 4$~T.

The comparison between theory and experiment shows a remarkable good agreement for both the diamagnetic shift and the Zeeman splitting.
In particular, we see that including band-mixing effects through the off-diagonal terms of the Luttinger Hamiltonian is crucial in order to reproduce the experimental results. In fact, calculations that neglect band mixing significantly overestimate the diamagnetic shift of all exciton states (Fig.~\ref{fig:diamagnetic-shift}) because they underestimate the exciton binding energy, as already shown at zero magnetic field in Fig.~\ref{fig:non-hydrogenic-series}.
Furthermore, due to the unequal band mixing in the $+$ and $-$ sectors (see Fig.~\ref{fig:ws-RydbergX}), we observe a nonlinear Zeeman splitting (Fig.~\ref{fig:zeeman}), where
the degree and character of the mixing vary across the $hh,ns$ series: The splitting of the $hh,1s$ exciton is strongly suppressed, whereas the $hh,ns$ Rydberg excitons exhibit a larger Zeeman splitting. Significantly, we observe a strong qualitative difference between the heavy-hole and light-hole splitting, which have opposite signs.
Conversely, if one neglects off-diagonal terms in the Luttinger Hamiltonian, then the Zeeman splitting varies linearly with the magnetic field and is identical for ground and excited states. This is in stark contrast to the experimental results, again highlighting the need for our multiband Luttinger model.

We stress that, in arriving at these results, we have fixed the microscopic model parameters to literature values (see Table~\ref{tab:parameters-model}), such that none were chosen to conveniently fit the experiments. However, although the 12 QWs have a nominal width of $d_z=20$~nm, we find that the agreement between experiment and theory is optimized by using an effective width $d_z=18$~nm in the model. Such a reduction in width of $\sim 2$~nm can be attributed to an off-center placement of the sample during growth, since the layer thickness at the center of the deposition region is slightly larger than towards the edges. We emphasize that this deviation does not compromise the uniformity of the quantum well width within the probed region.
In addition, we apply a rigid blueshift of $5.7$~meV to the theoretical diamagnetic shift in order to match the experimental value of the ground state at zero magnetic field. This energy offset may originate from unintentional doping of the sample. Additional states measured in the reflectance spectra (not shown), consistent with singlet and triplet trion resonances, support this interpretation (see the Supplemental Material of Ref.~\cite{short-paper}). The results for the Zeeman splitting are clearly not affected by this rigid shift.

\section{Conclusions}
\label{sec:conclusions}
To conclude, we have investigated the properties of ground and excited Rydberg exciton states in GaAs/Al$_{0.4}$Ga$_{0.6}$As quantum wells of $20$~nm width exposed to a perpendicular magnetic field up to 9~T, combining theory and experiment.
We have shown that it is essential to include heavy- and light-hole valence-band mixing effects in order to correctly describe the experimental results. Specifically, we have employed the multiband Luttinger model originally developed by Bauer and Ando~\cite{Bauer_AndoPRB1988} to describe magnetoexcitons in GaAs QWs, and developed an efficient numerical scheme to compute the eigenstates. We have demonstrated excellent agreement between theory and experiment for the diamagnetic shifts and Zeeman splittings of the first four heavy-hole Rydberg states and for the light-hole ground state, with the effective QW width the only adjustable theory parameter.

Our results show that band-mixing significantly increases the exciton binding energy, in turn reducing the diamagnetic shift compared to the case without hybridization.
Band mixing is also key to understanding the nonlinear Zeeman splittings, which we find have opposite signs for the heavy- and light-hold exciton states, and whose magnitude is much larger for excited heavy-hole Rydberg states compared with the ground state.
We are able to quantify the degree of hybridization of the exciton states and demonstrate that it increases with the magnetic field and is more pronounced for higher excited states. In particular, at low magnetic fields, the heavy-hole Rydberg states can still be regarded as quasi-hydrogenic excitations weakly perturbed by valence-band mixing effects. However, for larger magnetic fields, the excited states acquire a pronounced multiband character. 

Understanding the magnetic-field response of Rydberg excitons is of fundamental importance. A magnetic field tends to enhance the exciton oscillator strength and thus facilitates strong~\cite{Berger_PRB1996,Pietka_PRB2015,Pietka-Potemski_PRB2017} and very strong coupling~\cite{Brodbeck_PRL2017,Laird_PRB2022} to light, which can in turn fundamentally modify the properties of matter excitations~\cite{Khurgin_SSS2001,Zhang-Yamamoto_PRB2013,Yang_NJP2015}.
Our results are closely related to the joint accompanying paper~\cite{short-paper}, where we investigate the same heterostructure and focus on the very strong light–matter coupling regime enabled by patterning the slab with a one dimensional grating to access its waveguide modes. In that work, we show that heavy- and light-hole excitons hybridize into a single polariton mode and analyze how this very strongly hybridized state evolves with magnetic field.

An interesting perspective of our work is the possibility to model and study trion properties in quantum wells~\cite{Bar-Joseph_ReviewSST2005}, including their excited states~\cite{Jain-Glazov_PRL2025}. Our formalism, combined with the efficient numerical procedure that we have implemented, enables the exploration of the hybridization of heavy- and light-hole degrees of freedom within charged excitonic complexes—an area that remains largely unexplored.


\begin{acknowledgments}
We thank C. Tejedor for bringing to our attention the seminal work of Bauer and Ando~\cite{Bauer_AndoPRB1988}, as well as other key references before we even started this project.  DDFP acknowledges
financial support from FPU Grant No.~FPU24/03084. DDFP and FMM acknowledge financial
support from the Spanish Ministry of Science, Innovation and
Universities through the ``Maria de Maetzu'' Programme for Units of
Excellence in R\&D (CEX2023-001316-M) and from the Comunidad de Madrid
and the Spanish State through the Recovery, Transformation and
Resilience Plan [``Materiales disruptivos bidimensionales 2D'' (MAD2D-
  CM)-UAM7], as well as from the Ministry of Science, Innovation and
Universities MCIN/AEI/10.13039/501100011033, FEDER UE, project
No.~PID2023-150420NB-C31 (Q). MMP is supported through Australian
Research Council Future Fellowship FT200100619, and JL through
Australian Research Council Discovery Project DP240100569. JB and DB acknowledge financial support under the National Recovery and Resilience Plan (NRRP) under the PRIN grant 2022, funded by the European Union– NextGenerationEU– Project Title PENNA- CUP B53D23003790006- Grant Assignment Decree No. 957 by the Italian Ministry of University and Research (MUR). JB, DB, and DS acknowledge financial support from “Quantum Optical Networks based on Exciton polaritons” (Q-ONE, N. 101115575, HORIZON- EIC- 2022- PATHFINDER CHALLENGES EU project), ”Neuromorphic Polariton Accelerator” (PolArt, N.101130304, Horizon-EIC-2023Pathfinder Open EU project), “National Quantum Science and Technology Institute” (NQSTI, N. PE0000023, PNRR MUR project), “Integrated Infrastructure Initiative in Photonic and Quantum Sciences” (I-PHOQS, N. IR0000016, PNRR MUR project). Views and opinions expressed are, however, those of the authors only and do not necessarily reflect those of the European Union or European Innovation Council and SMEs Executive Agency (EISMEA). Neither the European Union nor the granting authority can be held responsible for them.
\end{acknowledgments}
\appendix

\section{Finite-difference scheme}
\label{App:finite-diff}
In this section, we describe the essential steps for the numerical solution of the coupled \sch equations~\eqref{eq:X-final-eqs} for the exciton envelope wavefunction after projection onto the basis of confined states as defined in Eq.~\eqref{eq:expansion-basis-confined}. 
The projection leads to a set of $N_e(2N_{hh}+2N_{lh})$ coupled equations for $\varphi^\alpha_{J_h k \ell n_e n_h}$, where $N_{e,hh,lh}$ is the number of confined states included in the expansion~\eqref{eq:expansion-basis-confined}. To simplify the notation, we represent this wave function vectorially as $\varphi^\alpha_{J_h k \ell n_e n_h} \mapsto \vegr{\varphi}_k$, retaining only the dependence on the magnitude of the momentum. The resulting eigenvalue problem can be written in matrix form as:
\begin{equation}
\label{eq:hamiltonian-projected-confined-basis}
    \mathbb{H}_{k} \bm{\varphi}_{k} + \int \frac{dk' k'}{2\pi} \mathbb{V}_{kk'} \bm{\varphi}_{k'} = E\bm{\varphi}_{k}\, .
\end{equation}
The matrix elements of the Coulomb potential, $\mathbb{V}_{kk'}$, diverge as $k \to k'$, and we regularize this divergence using the Land\'e subtraction scheme described in App.~\ref{App:subs-scheme}.  

We discretize $k$ on a grid $\{k_{i=1,...,N_k}\}$ and denote $\bm{\varphi}_i = \bm{\varphi}_{k_{i}}$. To do that, since $k\in[0,\infty)$, we consider the change of variable $k=\tan\beta$, with $\beta\in[0,\frac{\pi}{2})$. The $\beta$ interval is then discretized using the Gauss-Legendre quadrature, so that
\begin{align}
    k_i=&\tan \beta_i &
    dk_{i}=&\frac{d\beta_i}{\cos^2\beta_i}\, .
\end{align}

We define an auxiliary set of functions:
\begin{equation}
    \bm{\eta}_{k}=\partial_k \bm{\varphi}_{k}\, .
\end{equation}
First derivatives are evaluated using a midpoint finite-difference scheme\footnote{The same approach, on a linear grid, has been employed to efficiently find the energies of magnetoexcitons in TMD monolayers~\cite{Stier-Crooker_PRL2018}.} 
\begin{equation}
\label{eq:midpoint-deriv}
    \partial_k\bm{\varphi}_{k}=\frac{\bm{\varphi}_{i+1}-\bm{\varphi}_{i}}{k_{i+1}-k_i}=\frac{\bm{\eta}_{i}+\bm{\eta}_{{i+1}}}{2}\, ,
\end{equation}
with $i=1,...,N_k-1$,
while second derivatives are
\begin{align}
    \partial_k^2\bm{\varphi}_{k}=&\frac{\bm{\eta}_{{i+1}}-\bm{\eta}_{{i}}}{k_{i+1}-k_{i}}\, .
\end{align}
The other terms in the system of equations~\eqref{eq:hamiltonian-projected-confined-basis} are also evaluated on the grid midpoints; for example, the Coulomb term reads
\begin{equation*}
    \int \frac{dk' k'}{2\pi}\mathbb{V}_{k,k'}\bm{\varphi}_{k'}
    = \sum_j D_j^2\mathbb{V}_{\frac{k_i+k_{i+1}}{2},\frac{k_j+k_{j+1}}{2}}\frac{\bm{\varphi}_{j}+\bm{\varphi}_{{j+1}}}{2}\, ,
\end{equation*}
where $D_j^2=\frac{k_j+k_{j+1}}{2}\frac{dk_j + dk_{j+1}}{4\pi}$. 
Thus, we end up with the following system of $N_e(2N_{hh} + 2N_{lh}) (2N_k-2)$ coupled equations:
\begin{subequations}
\label{eq:2-discre}
\begin{align}    
\label{eq:discretized-hamil-in-k}
    &\left[\mathbb{H}^{\varphi}_{ ij} \mathbb{M}^{}_{jl}\vegr{\varphi}_{l}+\mathbb{H}^{\eta}_{ij} \mathbb{M}^{} _{jl}\vegr{\eta}_{l} \notag\right.\\
    &\left. + D_j^2\mathbb{V}_{\frac{k_i+k_{i+1}}{2},\frac{k_j+k_{j+1}}{2}} \mathbb{M}^{}_{jl}\vegr{\varphi}_{l}\right] = \mathbb{M}_{il}\vegr{\varphi}_{l} \\
    \label{eq:def-mid-point-deriv}
    &\mathbb{M}_{ij}\vegr{\eta}_{j} - \frac{\vegr{\varphi}_{i+1}-\vegr{\varphi}_{i+1}}{k_{i+1}-k_i}=0\, ,
\end{align}    
\end{subequations} 
where repeated indices are implicitly summed and where $\mathbb{M}_{ij}=\frac{1}{2}(\delta_{i,j}+\delta_{i,j+1})$. 
The missing $N_e(2N_{hh} + 2N_{lh}) 2$ equations to match the number of unknowns are given by the following boundary conditions (see Ref.~\cite{Stier-Crooker_PRL2018} for the $\ell=0$ case):
\begin{subequations}
\begin{align}
    \vegr{\varphi}_{N_k-1} &= 0 & \vegr{\eta}_{1} &= 0 & \text{for}\  \ell&=0\\
    \vegr{\eta}_{N_k-1} &= 0 & \vegr{\varphi}_{1} &= 0 & \text{for}\  \ell&\ne 0\, 
\end{align}
\end{subequations}
The two sets of coupled equations~\eqref{eq:2-discre} can be reduced to one set by solving explicitly~\eqref{eq:def-mid-point-deriv} for $\vegr{\eta}_i$ and substituting into~\eqref{eq:discretized-hamil-in-k}. We then solve the problem by direct diagonalization.

\section{Subtraction scheme for the Coulomb potential in finite-width  QWs}
\label{App:subs-scheme}
In this appendix, we describe the Land\'e subtraction scheme used to regularize the pole of the Coulomb potential $V_{kk'\ell}(z)$~\eqref{eq:FT-Coulomb-term-lz} at $k=k'$. This approach generalizes schemes previously developed for 2D excitons~\cite{Laird_PRB2022,deLaFuentePico_PRB2025} and 2D trions~\cite{Kumar-Levinsen_PRB2025} to QWs of finite width.
We define the auxiliary function
\begin{equation}
\label{eq:g-subs-def}
    g_{kk'\ell} (z) = \frac{2 k^2}{k^2+k'^2} V_{kk'\ell}(z) \, ,
\end{equation}
which allows us to rewrite the Coulomb integrals appearing in the \sch equations~\eqref{eq:X-final-eqs} as follows:
 \begin{multline}
 \label{eq:subs-scheme}
    \int_0^\infty \frac{dk'k'}{2\pi} V_{k k' \ell}(z_e-z_h) \varphi_{J_h k' \ell}^{\alpha}(z_e,z_h)\\
    = \mathcal{K}_{k\ell}(z_e-z_h) \varphi_{J_h k \ell}^{\alpha}(z_e,z_h) \\ 
    +\int_0^{\infty} \frac{dk' k'}{2\pi} \left[V_{k k' \ell}(z_e-z_h) \varphi_{J_h k' \ell}^{\alpha}(z_e,z_h)\right.\\
    \left. - g_{kk'\ell} (z_e-z_h) \varphi_{J_h k \ell}^{\alpha}(z_e,z_h)\right]\, ,     
 \end{multline}
 where 
\begin{multline}
\label{eq:K-function}
     \mathcal{K}_{k\ell}(z) = \int\frac{dk' k'}{2\pi} g_{kk'\ell}(z) \\
     =-\frac{2\pi e^2}{\varepsilon^2} \int\frac{dk' k' d\theta}{(2\pi)^2} \frac{2 k^2}{k^2+k'^2}  \cos(\ell \theta)\frac{e^{-|z|Q_{kk'}(\theta)}}{Q_{kk'}(\theta)} \, ,
\end{multline}
with $Q_{kk'}(\theta)=\sqrt{k^2+k'^2-2 k k' \cos\theta}$.
The prefactor in~\eqref{eq:g-subs-def} has been chosen~\cite{Laird_PRB2022} so that $g_{kk\ell} (z) = V_{kk\ell} (z)$ and to speed up numerical convergence when $k' \to \infty$.

The integral in~\eqref{eq:K-function} is convergent and, for $z=0$ (2D limit), it can be evaluated analytically. For $z\ne 0$, the angular integration over $\theta$ can be conveniently replaced by a smooth Bessel-kernel integral through the Hankel expansion, 
which reads as:
\begin{equation}
   \frac{e^{-q |z|}}{q}=\int_0^{\infty} dr \frac{r J_0(q r)}{\sqrt{r^2+|z|^2}} \, .
\end{equation}
The above expansion, together with Graf's addition theorem~\cite{WatsonBessel_Book},
\begin{equation}
    J_0(Q_{kk'}(\theta)r)=\sum_{m=-\infty}^{\infty} J_{m}(kr)J_{m}(k'r)e^{im\theta}\, ,
\end{equation}
allows us to perform the integration over $\theta$. The integration over $k'$ is also performed analytically. This enables us to recast the integral 
~\eqref{eq:K-function} in the following form:
\begin{equation}
    \mathcal{K}_{k\ell}(z) = k^2\int_0^{\infty} dr \frac{r}{\sqrt{r^2+|z|^2}} J_{\ell}(kr)f_{\ell}(kr)\, ,
\end{equation}
where $J_{\ell} (x)$ are the Bessel function of the first kind of order $\ell$ and $f_{\ell}(x)$ are 
functions which decay at least as fast as $1/x^2$. 
Note that the original 2D integral~\eqref{eq:K-function} over $k', \theta$  is reduced to a 1D integral over $r$. Moreover, numerical convergence is improved by removing the ill-conditioned term $Q_{kk'}(\theta)^{-1}$ which diverges at $k=k', \theta=0,2\pi$. The explicit expressions of $ f_\ell(kr)$ for $\ell=0,1,2,3$ are:
\begin{align}
    f_0(kr)=&2K_0(kr)\\
    f_1(kr)=&\pi M_1(kr)+2\\
    f_2(kr)=&\frac{4}{k^2r^2}-2K_2(kr)\\
    f_3(kr)=&\frac{2+3\pi(I_3(kr)-L_1(kr))+\frac{12\pi}{kr}L_2(kr)}{3}\,,
\end{align}
where $I_n(x)$ and $K_n(x)$ are the modified Bessel functions of the first and second kind respectively, of order $n$, and $L_n(x)$ is the modified Struve function of order $n$.

 \begin{figure}
\centering
\includegraphics[width=1.0\linewidth]{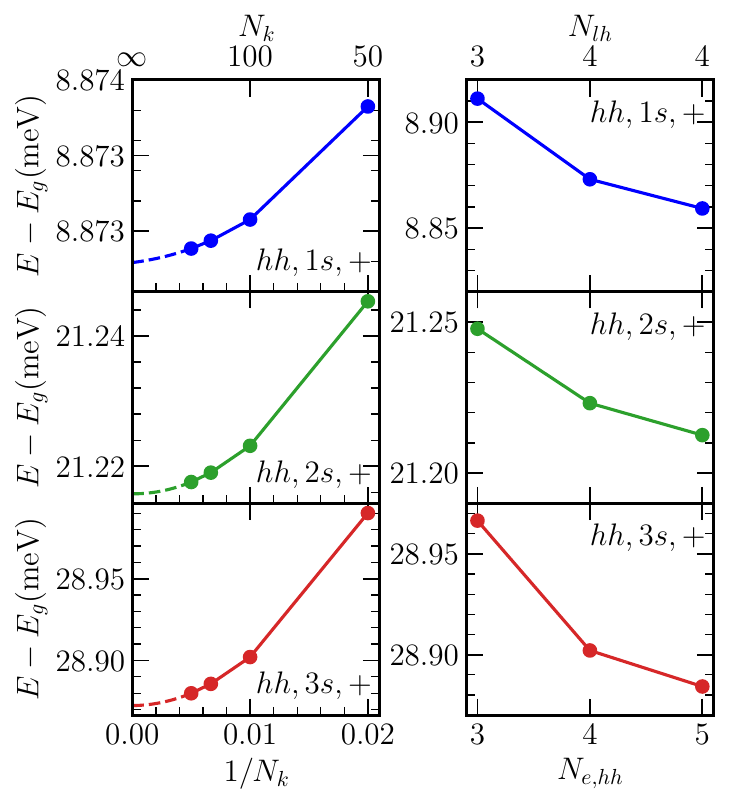}
    \caption{Convergence of the energies of the $ns$ exciton states in the $hh+$ sector at $B=3.14$~T as a function of the number of $k$-grid points $N_k$ (left panel, for $N=N_{e,lh,hh}=4$) and of the number of confined states $N_{e,lh,hh}$ (right panel, for $N_k=100$). $N_{lh}$ is limited to $4$ because it is the maximum number of QW confined $lh$ states that are bound. The $N_k \to \infty$ limits are obtained by cubic extrapolation (dashed lines). The QW width is $d_z=18$~nm.}
\label{fig:convergence-Nk}
\end{figure}
\begin{figure}
    \centering
    \includegraphics[width=\linewidth]{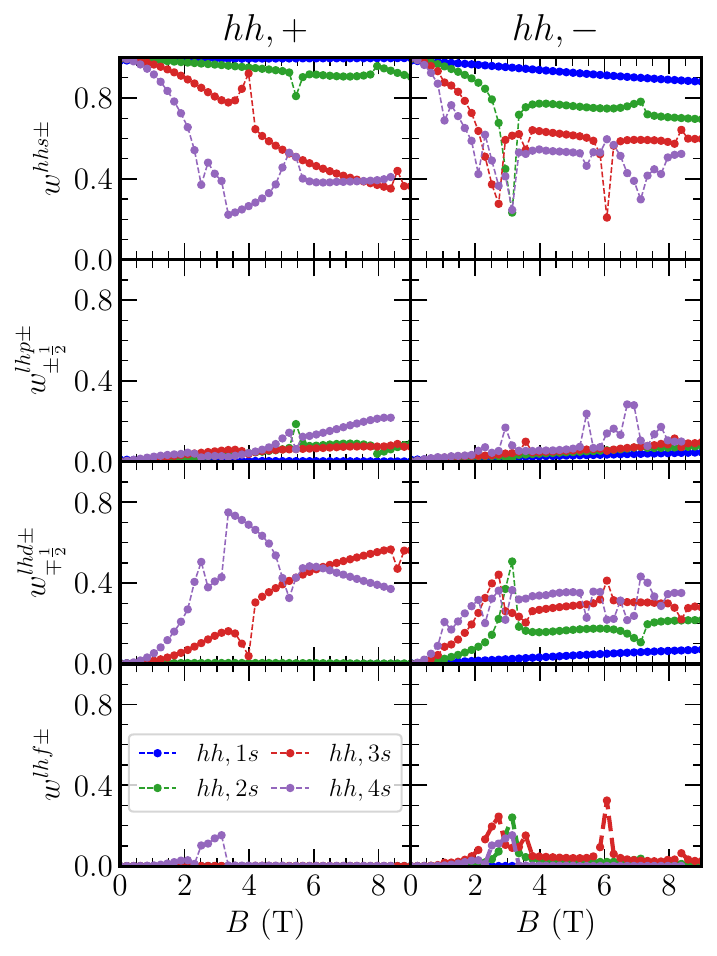}
    \caption{$s$-wave, $d$-wave, $p$-wave, and $f$-wave fractions of the first $hh,ns$ exciton states of the $hh,\pm$ sectors---see Fig.~\ref{fig:Xenergies-colored-osc-str}.
    }
\label{fig:rydberg-fractions-orbital-components}
\end{figure}
%
\section{Numerical convergence}
\label{app:convergence}
Having illustrated in the previous sections the finite-difference and subtraction schemes used to numerically solve the exciton equations~\eqref{eq:X-final-eqs}, we now briefly examine the convergence of our results with respect to the discretization parameters, i.e., the dependence on the number of grid points in $k$, $N_k$, and the number of confined states~\eqref{eq:expansion-basis-confined}, $N_{e,hh,lh}$. 
In Fig.~\ref{fig:convergence-Nk} we show the dependence of the $hh+$ exciton energies on $N_k$ and $N_{e,hh,lh}$ for a fixed magnetic field. Extrapolation to the $N_k \to \infty$ limit indicates that, relative to the values obtained at $N_k=200$, the error is below $10^{-4}$~meV for the $1s$ state and below $10^{-2}$~meV for the $2s$ and $3s$ states. 

For all the numerical results shown in the main text at QW width $d_z = 18$~nm, we use $N_k = 200$ and $N_{e,hh,lh} = 4$. For smaller widths, the number of confined states decreases: fewer than four $lh$ states exist for $d_z \leq 14$~nm, fewer than four electron states for $d_z \leq 12$~nm, and fewer than four $hh$ states for $d_z = 6$~nm. Accordingly, in Fig.~\ref{fig:Energy-LZdependence} we take $N_{e,hh,lh}$ to be the maximum number of bound states available for each $d_z$.
Since Fig.~\ref{fig:Energy-LZdependence} is restricted to $B=0$, the $hh,ns$ states associated with the subband $(n_e=0,n_h=0)$ are not expected to hybridize with barrier-edge states with $\epsilon_{n_e} > V_c$ and $\varepsilon_{J_hn_h} > V_v$. Therefore, the results obtained from our truncated expansion are expected to remain reliable.

%
\section{Orbital momentum fractions of $hh$ excitons}
\label{app:Rydberg-comp}
Similar to the definition of the total $s$-wave fraction~\eqref{eq:s-wave-fraction}, we can define the higher orbital angular momentum fractions:
\begin{subequations}
\begin{align}
 w^{lh p\pm}_{\pm\frac{1}{2}}=&\sum_{n_e,n_h}\int \frac{dk k}{2\pi} \left|\varphi^{(Jz=\pm1 J_e=\mp \frac{1}{2})}_{J_h=\pm\frac{1}{2} k \ell=\pm1 n_e n_{h}} \right|^2\\
 w^{lhd\pm}_{\mp\frac{1}{2}}=&\sum_{n_e,n_h}\int \frac{dk k}{2\pi} \left|\varphi^{(Jz=\pm1 J_e=\mp \frac{1}{2})}_{J_h=\mp\frac{1}{2} k \ell=\pm2 n_e n_{h}} \right|^2\\
    w^{hh f \pm}=&\sum_{n_e,n_h}\int \frac{dk k}{2\pi}|\varphi^{(J_z=\pm 1 J_e=\mp \frac{1}{2})}_{J_h=\mp\frac{3}{2} k\ell=\pm 3 n_en_h}|^2 \, .
\end{align}
\end{subequations}
Note that, because the $lh, p$ and $lh,d$ states appear in both the $hh$ and $lh$ sectors for a given $J_z$ (see Table~\ref{tab:ell-Jh-quartets}), the corresponding value of $J_h$ must be specified in the definition. We plot in Fig.~\ref{fig:rydberg-fractions-orbital-components} the $p$-, $d$-, and $f$-wave fractions of the excitons in the $hh \pm$ sectors---we repeat also the $s$-wave fraction of Fig.~\ref{fig:ws-RydbergX} for comparison. 
As noted in the main text, mixing increases with magnetic field and is more pronounced for higher excited states. The $hh,ns,-$ states mix predominantly with $lh,d,-$ states and, for $n>1$, acquire a small $hh,f,-$ component in the range $2 \lesssim B \lesssim 4$~T. In contrast, the $hh,1s,+$ state retains an almost purely $hh$ $s$-wave character, with $w^{hh s} \simeq 1$ over the entire field range. The $hh,2s,+$ state mixes weakly with $lh,p,+$ states, while the $hh,3s,+$ and $hh,4s,+$ states mix mainly with $lh,d,+$ states.

%

\end{document}